\documentclass[amsmath,amssymb,12pt]{revtex4-1}
\usepackage[utf8]{inputenc} 
\usepackage{graphicx}
\usepackage{bbold}
\usepackage{float}
\usepackage{natbib}
\usepackage{wrapfig}
\usepackage[english]{babel}
\usepackage[normalem]{ulem}
\newcommand{\del}{\partial}

\begin{document}

	\title{Ott-Antonsen ansatz for the D-dimensional Kuramoto model: a constructive approach.}

	\author{Ana Elisa D.  Barioni}

	\author{Marcus A. M. de Aguiar }
	
	\affiliation{Instituto de F\'isica `Gleb Wataghin', Universidade Estadual de Campinas, Unicamp 13083-970, Campinas, SP, Brazil}
	
	\begin{abstract}
		
		Kuramoto's original model describes the dynamics and synchronization behavior of a set of interacting oscillators represented by their phases. The system can also be pictured as a set of particles moving on a circle in two dimensions, which allows a direct generalization to particles moving on the surface of higher dimensional spheres. One of the key features of the 2D system is the presence of a continuous phase transition to synchronization as the coupling intensity increases. Ott and Antonsen proposed an ansatz for the distribution of oscillators that allowed them to describe the dynamics of the transition's order parameter with a single differential equation. A similar ansatz was later proposed for the D-dimensional model by using the same functional form of the 2D ansatz and adjusting its parameters. In this paper we develop a constructive method to find the ansatz, similarly to the procedure used in 2D. The method is based on our previous work for the 3D Kuramoto model where the ansatz was constructed using the spherical harmonics decomposition of the distribution function. In the case of motion in a D-dimensional sphere the ansatz is based on the hyperspherical harmonics decomposition. Our result differs from the previously proposed ansatz and provides a simpler and more direct connection between the order parameter and the ansatz.

	\end{abstract}
	
	\maketitle

	\section{Introduction}
	
	The phenomenon of synchronization in systems of coupled oscillators is a subject of intense study and increasing importance. It has been found as a crucial aspect in many fields including biological, technological and physical systems, such as the synchronization behavior of groups of cardiac pacemaker cells \cite{Osaka2017}, coupled metronomes, large groups of fireflies \cite{Ermentrout1991}, \cite{Buck1968}, biochemical oscillators \cite{Kiss1676}, oscillating neutrinos \cite{Pantaleone1998} and neuronal synchronization \cite{Chandra2017}.  In neural systems, the study of synchronization is related to brain rhythms an brain physiology, pathology, and cognition. In this context, the collective oscillation is related with negative phenomena, such as epilepsy and tremor activity, but it is also related with brain rhythms and cognition \cite{Guevara2017}. The large number and wide variety of applications has motivated the study of the mathematical methods that describes the global behavior of large systems and also in higher dimensions.  The two-dimensional model originally proposed by Kuramoto describes a set of N coupled oscillators by their phases $\theta_i$ and natural frequencies $\omega_i$, chosen from a symmetric distribution $g(\omega)$. The equations that determine the dynamics are
	\begin{equation}
		\frac{d\theta_i}{dt}= \omega_i+\frac{K}{N}\sum_{j=1}^{N}\sin(\theta_j-\theta_i), 
	\end{equation}
	where K is the coupling strength between the oscillators. In order to determine the degree of synchronization an order parameter is defined as 
	\begin{equation}
		z = p e^{i \psi} \equiv \frac{1}{N} \sum_{i=1}^N e^{i\theta_i},
		\label{paraord}
	\end{equation}
	where the case with $p=0$ indicates disordered motion and $p=1$ indicates full synchrony. The order parameter measures the collective behavior of the system and shows activity patterns and the effects of the interactions between each pair of dynamical agents. An important step towards understanding the system analytically was made by Ott and Antonsen \cite{Ott2008}, who  proposed a method to calculate the dynamics of the order parameter of a large set of coupled particles involving only two time dependent parameters. They considered the limit where $N\rightarrow\infty$, so that the oscillators are distributed in the circle according to a density function. Then, by the conservation of the number of oscillators, the continuity equation is satisfied and the authors proposed an ansatz for the distribution of the oscillators on the unit circle. The ansatz parameters are connected to the order parameter by the distribution of natural frequencies and the model is reduced to two equations of motion. 
	
	In the context of many applications, it is important to consider synchronization in higher-dimensional spaces. For example, three dimensional brain networks and flocks moving in three and higher-dimensional space  \cite{Olfati-Saber2006}. The Kuramoto model was also extended to any number of dimensions and shown to have discontinuous phase transitions in odd dimensions \cite{chandra2019continuous}.  The 2D ansatz was extended to the D-dimensional model by using the same functional form of the 2D ansatz and adjusting its parameters in Chandra et al. \cite{chandra2019complexity}. In a previous work \cite{BarioniAguiar2021}, we took a different approach and constructed a new ansatz for the 3 dimensional case based on spherical harmonics decomposition of the distribution function. We derived the phase diagram of equilibrium solutions for several distributions of natural frequencies and found excellent agreement with numerical solutions for the full system dynamics.  Our ansatz differs from that obtained by Chandra et al. \cite{chandra2019complexity}. Although the vector equation satisfied by the ansatz parameter $\vec{\rho}$ is the same as obtained in \cite{chandra2019complexity},  the connection with the order parameter $\vec{p}$  is simpler in our approach. The relationship we find is the natural extension of the 2D case, where $\vec{p}$ is the integral of the order parameter $\vec{\rho}$ over $G(\vec{\omega})$.
	\\
	\indent In this paper we aim to extend the ansatz we proposed for 3 dimensions to any dimension. Following \cite{chandra2019continuous} we write the Kuramoto equations in vector form, with each oscillator being represented as an unit vector rotating on the surface of a hypersphere $S_{D-1}$. We solve the continuity equation by expanding the distribution of oscillators in hyperspherical harmonics and making an ansatz for the expansion coefficients. Then we derive the equations of motion for the ansatz parameters and compare the results of our analytical treatment with numerical simulations. The paper is organized as follows: in section II, we present the D-dimensional vector formulation of the Kuramoto model and establish the spherical coordinates in  D-dimensions. In section III, we write the continuity equation, propose the ansatz for the distribution of oscillators and derive the equations of motion for the ansatz parameters. In Appendix \ref{app1} we review the formalism of the spherical harmonics in higher dimensions and state and demonstrate some properties that will be important for the development of the theory.

	\section{The D-dimensional Kuramoto Model}
	
	\subsection{Vector formulation of the Kuramoto Model}
	The D-dimensional Kuramoto model is a direct extension of the original 2D model and represents unit vectors $\vec{\sigma_i}$ rotating on the surface of a hypersphere $S_{D-1}$. The equations are given by \cite{chandra2019continuous}
	\begin{equation}
		\frac{d \vec{\sigma_i}}{d t} = \mathbf{W}_i \vec{\sigma_i} + \frac{1}{N}  
		\sum_j [\mathbf{K}_i \vec{\sigma_j} - (\vec{\sigma_i}\cdot \mathbf{K}_i \vec{\sigma_j}) \vec{\sigma_i}],
		\label{3dkura}
	\end{equation}
	where  $\mathbf{W}_i$ is a $D \times D$ anti-symmetric matrix, with $D(D-1)/2$ independent frequencies for each oscillator, and $\mathbf{K}_i$ is the $D \times D$ coupling matrix, whose elements might depend on $\vec{\sigma_i}$. The natural frequencies of each oscillator are drawn from a normalized distribution $G(\mathbf{W})$.
	
	The order parameter, which measures the degree of phase synchronization of the N oscillators system, is given by
	\begin{equation}
		\vec{p} = \frac{1}{N}\sum_i \vec{\sigma_i}.
		\label{vecpar}
	\end{equation}
	To simplify the derivations we define an auxiliary vector $\vec{q}_i = {\mathbf K}_i \, \vec{p}$ so that  the Kuramoto equations can be rewritten as 
	\begin{equation}
		\frac{d \vec{\sigma_i}}{d t} = \mathbf{W}_i \vec{\sigma_i} +  [ \vec{q}_i - (\vec{\sigma_i}\cdot \vec{q}_i) \vec{\sigma_i}].
		\label{kuramotogenk}
	\end{equation}

	\subsection{Spherical coordinates in D dimensions}
	
	Spherical coordinates in a D-dimensional space are analogous to the 3 dimensional ones, with one radial component  and $D-1$ angular components $\{\theta_1, ..., \theta_{D-1}\}$, where $\theta_2, ..., \theta_{D-1}$ range over $[0,\pi]$ and $\theta_1$ ranges over $[0,2\pi)$. This extends the definition of the polar and azimuth angles in 3D, with $\theta\rightarrow\theta_2$ and $\phi\rightarrow\theta_1$.
	
	Cartesian coordinates $x_i$ can be computed in terms of the spherical parameters as
	\begin{equation}
		\begin{split}
			x_1 &= r\sin{\theta_{D-1}}\ldots\sin{\theta_2}\cos{\theta_1}\\
			x_2 &= r\sin{\theta_{D-1}}\ldots\sin{\theta_2}\sin{\theta_1}\\
			\vdots\\
			x_{D-2} &=r\sin{\theta_{D-1}}\sin{\theta_{D-2}}\cos{\theta_{D-3}}\\
			x_{D-1} &=r\sin{\theta_{D-1}}\cos{\theta_{D-2}}\\
			x_{D} &=r\cos{\theta_{D-1}}.
		\end{split}
	\end{equation}
	
	The spherical surface area element (or generalized solid angle) is given by the determinant of the Jacobian matrix with unitary radius, which leads to 
	\begin{equation}
		d^D S= \sin^{D-2}\theta_{D-1}\sin^{D-3}\theta_{D-2}\ldots\sin{\theta_2} d\theta_{D-1}\ldots d\theta_1.
		\label{VolEl}
	\end{equation}

	
	\section{Dynamics of the order parameter}

	\subsection{Continuity equation}
	
	In the limit where the number of oscillators $N \rightarrow \infty$ it is convenient to define a distribution function that describes, at time t, the state of the oscillator system. The function $f(\textbf{W},\vec \sigma, t)$ represents the density of oscillators with natural frequencies specified by  $\textbf{W}$ and position given by the unit vector $\vec\sigma=\vec\sigma(\theta_1, \ldots, \theta_{D-1})$ at time $t$. The following normalization conditions must then be satisfied
	\begin{equation}
		\int \int_\mathcal{S} f(\textbf{W}, \vec\sigma, t) d\textbf{W}d^DS = 1, 
	\end{equation}
	and
	\begin{equation}
		\int_\mathcal{S} f(\textbf{W}, \vec\sigma, t) d\textbf{W} = G(\textbf{W}), 
	\end{equation}
	where $G(\textbf{W})$ is the distribution of anti-symmetric matrices of natural frequencies. In terms of $f$ the order parameter becomes
	\begin{equation}
		\vec{p}(t) = \int \int_\mathcal{S} \vec\sigma f(\textbf{W}, \vec\sigma, t) d\textbf{W}d^DS.
		\label{param}
	\end{equation}
	
	Conservation of the number of oscillators leads to the continuity equation
	\begin{equation}
		\partial_t f +\vec\nabla\cdot[\vec{v}f]=0,
	\end{equation}
	where the velocity field is given by $\vec{v} = \mathbf{W} \vec{\sigma} +  [ \vec{q} - (\vec{\sigma}\cdot \vec{q}) \vec{\sigma}].$

	Following Appendix B of \cite{chandra2019continuous} we rewrite the continuity equation in a more suitable way as
	\begin{equation}
		\frac{\del f}{\del t} + [\vec\nabla_\mathcal{S}f(\vec\sigma, \textbf{W}, t)-(D-1)f(\vec\sigma, \textbf{W}, t)\vec\sigma]\cdot\textbf{K}\vec p + \textbf{W}\vec\sigma\cdot\vec\nabla_\mathcal{S}f(\vec\sigma, \textbf{W}, t)=0.
		\label{ContinuityEq}
	\end{equation}

	
	\subsection{Ansatz for the density function}
	
	Ott and Antonsen \cite{Ott2008} proposed an ansatz for density function of the 2D Kuramoto model that allowed them to reduce the dynamics of infinitely many oscillators to a simple equation for the order parameter. 
	They expanded the density function in Fourier series and made a simple but consistent choice for the coefficients. The same idea was applied to the 3D model in  \cite{BarioniAguiar2021} where the density 
	function was expanded in spherical harmonics. Here we generalize the procedure for D-dimensions, expanding the
	density function in hyperspherical harmonics. The generalization of the spherical harmonics
	to D dimensions is given by
	
	\begin{equation}
		Y^{l_1}_{l_2,\ldots,l_{D-1}}(\theta_1, \ldots, \theta_{D-1})= N_{l_{D-1},l_{D-2}}\sin^{l_{D-2}}{
			\theta_{D-1}}C_{l_{D-1}-l_{D-2}}^{l_{D-2}+\frac{D}{2}-1}(\cos{ \theta_{D-1}})Y^{l_1}_{l_2,\ldots,l_{D-2}}(\theta_1, \ldots, \theta_{D-2}),
		\label{HSH}
	\end{equation}
	where $N_{l_{D-1},l_{D-2}}$ is the normalization constant 
	
	\begin{equation}
		N_{l_{D-1},l_{D-2}}=\sqrt{\frac{(l_{D-1}-l_{D-2})!(l_{D-1}+\frac{D}{2}-1)}{\pi\Gamma(l_{D-1}+l_{D-2} + D-2)}}2^{l_{D-2}+\frac{D-3}{2}}\Gamma(l_{D-2}+\frac{D}{2}-1) 
		\label{normconst}
	\end{equation}
	and $C^\alpha_n$ are the Gegenbauer polynomials. The development of this generalization can be found in Appendix \ref{app1}. Note that $l_1$ is the index associated with $\theta_1$, which ranges over $[0,2\pi)$ and is analogous to $\phi$ in the 3D coordinates. For that reason we identify $l_1\equiv m$.
	
	The density function $f$ can now be expanded as
	\begin{equation}
		f(\theta_1, \ldots, \theta_{D-1}, \textbf{W}, t)= G(\textbf{W})\sum_{l_{D-1}=0}^\infty\sum_{l_{D-2}=0}^{l_{D-1}} \ldots \sum_{m=-l_2}^{l_2} f^m_{l_2, \ldots, l_{D-1}}(\textbf{W}, t)Y^m_{l_2,\ldots,l_{D-1}}(\theta_1, \ldots, \theta_{D-1}).
		\label{expansion}
	\end{equation}
	
	In order to solve the continuity, the expansion of the density function is not sufficient, once it generates a set of coupled nonlinear differential equations for the coefficients. Following \cite{Ott2008,BarioniAguiar2021} we restrict our attention to a special class of functions
	\begin{equation}
		f^m_{l_2, \ldots, l_{D-1}} = \rho^{l_{D-1}}Y^{m *}_{l_2,\ldots,l_{D-1}}(\Theta_1, \ldots, \Theta_{D-1}) ,
		\label{ansatz}
	\end{equation}
	where we define the vector $\vec\rho=\rho \hat r(\Theta_1, \ldots, \Theta_{D-1})$ as the ansatz parameter. The exponent of $\rho$ is determined when the \textit{generalized addition theorem for hyperspherical harmonics} is applied to Eq.(\ref{expansion}), once the $\rho$ must be factored out.

	Using this ansatz we can simplify the density function expression using the hyperspherical harmonics  properties. Substituting Eq.(\ref{ansatz}) into Eq.(\ref{expansion}) we find 
	\begin{equation}
		\begin{split}
			f(\theta_1, \ldots, \theta_{D-1}, \textbf{W}, t)= G(\textbf{W})\sum_{l_{D-1}=0}^\infty\sum_{l_{D-2}=0}^{l_{D-2}} \ldots \sum_{m=-l_2}^{l_2} \rho^{l_{D-1}}Y^{m *}_{l_2,\ldots,l_{D-1}}(\Theta_1, \ldots, \Theta_{D-1}) \\
			\times Y^m_{l_2,\ldots,l_{D-1}}(\theta_1, \ldots, \theta_{D-1}).
			\label{ExpAnsatz}
		\end{split}
	\end{equation}

	Now, with the addition theorem formula (Eq.(\ref{AddTheorem})), we obtain
	\begin{equation}
		f(\theta_1, \ldots, \theta_{D-1}, \textbf{W}, t)= G(\textbf{W})\sum_{l_{D-1}=0}^\infty\left(\frac{2l_{D-1}}{n-2}+1\right)\frac{C^{\frac{D-2}{2}}_{l_{D-1}}(\hat\rho\cdot\hat r)}{\mathcal{A}_D},
	\end{equation}
	where $\hat{r}=\hat{r}(\theta_1, \ldots, \theta_{D-1})$ is the unit vector on the sphere. From the following Gegenbauer property 
	\begin{equation}
		\sum_{l_{D-1}=0}^\infty C^\alpha_{l_{D-1}}(x)y^{l_{D-1}}=\frac{1}{(1-2xy+y^2)^\alpha},
	\end{equation}
	we are led to the reduced expression of $f$
	\begin{equation}
		f(\theta_1, \ldots, \theta_{D-1}, \textbf{W}, t)=\frac{\Gamma(\frac{D}{2})}{2\pi^{\frac{D}{2}}} G(\textbf{W})\frac{1-\rho^2}{(1+\rho^2-2\rho\hat\rho\cdot\hat r)^{\frac{D}{2}}}.
		\label{distransatz}
	\end{equation}
	
	Notice that the term multiplying the distribution of natural frequencies is the related to the Poisson kernel for the unitary ball, $P(\vec\rho,\vec r)$ and the density can be written as
	\begin{equation}
		f =G(\textbf{W})P(\vec\rho,\vec r) .
		\label{ansatzPoisson}
	\end{equation}

	
	\subsection{Ansatz and order parameter connection}
	
	Using this ansatz and some properties of the hyperspherical harmonics, we can write the connection between the ansatz and order parameter in a very simple expression. 
	
	We show, in Appendix \ref{app1}, that the spherical coordinates in dimension D can be written in terms of the spherical harmonics with $l_{D-1}=1$. Then we write the order parameter in Eq. (\ref{param}) as 
	\begin{equation}
		\begin{split}
			\vec{p}(t) = \int \int_\mathcal{S} \vec\sigma \, G(\textbf{W})\sum_{l_{D-2}=0}^{1}...\sum_{m=-l_2}^{l_2} \rho^{\alpha(l_1, \ldots, l_{D-2}, 1)}Y^{m *}_{l_2,\ldots,l_{D-2}, 1}(\Theta_1, \ldots, \Theta_{D-1})\\
			\times Y^m_{l_2,\ldots,l_{D-1}}(\theta_1, \ldots, \theta_{D-1}) \, d\textbf{W}\,d^DS,
			\label{paramAnsatz}
		\end{split}
	\end{equation}
	where we used the orthogonality of the hyperspherical harmonics.
	
	Applying the addition theorem formula for $l_{D-1}=1$ in Eq.(\ref{paramAnsatz}), and with $C_1^\alpha(x)=2\alpha x$, we are led to
	\begin{equation}
		\vec p(t) = \int G(\textbf{W})\,\rho\,\frac{D}{\mathcal{A}_D}\left[\int \hat r\hat r^\top d^DS\right]\cdot\hat \rho\, d\textbf{W}\,.
		\label{paramAl}
	\end{equation}
	
	The angular integral of the dyadic matrix is given by
	\begin{equation}
		\int \hat r\hat r^\top d^DS = \frac{\mathcal{A}_D}{D}\textbf{1}.
		\label{dyadic}
	\end{equation}
	
	This result is proven in Appendix \ref{app2}. Then, substituting Eq.(\ref{dyadic}) into Eq.(\ref{paramAl}), the order parameter is
	\begin{equation}
		\vec p(t) = \int \vec\rho\,\, G(\textbf{W})\, d\textbf{W}.
		\label{paramFinal}
	\end{equation}
	
	Surprisingly, this is the same relation that we obtained for the 2D and 3D model, as we can see in \cite{BarioniAguiar2021}. The explicit D dependence cancels out, and we have a simple connection between the order and the ansatz parameters. The ansatz vector $\vec\rho(\omega, t)$ is interpreted as the order parameter of the subset of oscillators with natural frequency $\omega$.
	
	
	\subsection{Continuity equation for ansatz distribution}

	If we now substitute the expression of the density function in terms of the ansatz parameter, Eq. (\ref{distransatz}), in the continuity equation, Eq.(\ref{ContinuityEq}), we obtain
	\begin{equation}
		\begin{split}
			(1+\rho^2 -2\rho\hat r\cdot\rho)(&-2\vec\rho\cdot\del_t\vec\rho)-D(1-\rho^2)(\vec\rho\cdot\del_t\vec\rho - \hat r\cdot\del_t\vec\rho) +\\ (1-\rho^2)\{D(\vec\rho\cdot\textbf{K}\vec p)&+(D-2)(\vec\rho\cdot\hat r)(\textbf{K}\vec p\cdot\hat r) - (D-1)(\textbf{K}\vec p\cdot\hat r)(1+\rho^2)-D\hat r\cdot\textbf{W}\vec\rho\}=0.
			\label{continuity1}
		\end{split}
	\end{equation}
	
	Remembering that we defined $\textbf{K}\vec p= \vec q$, notice that the first two terms inside the curly brackets of Eq.(\ref{continuity1}) gives us
	\begin{equation}
		D\vec\rho\cdot\vec q + (D-2)p_rq_r .
		\label{CouplingTerms}
	\end{equation}
	
	Notice that the radial terms in (\ref{CouplingTerms}) have no counterpart in the continuity equation. As we did in \cite{BarioniAguiar2021}, this problem is solved by exploring the invariance of the exact equations of Kuramoto in the radial part of $\vec q$. We then choose the coupling vector $\vec q$ so that the undesired terms are canceled and the exact equations are not affected
	\begin{equation}
		\vec q =\left[\textbf{1}-\frac{D-2}{2(D-1)}\hat{r}\hat r^\top\right]\textbf{K}\vec p + \beta \hat r .
		\label{coupling}
	\end{equation}
	
	Applying this choice in the continuity Eq.(\ref{continuity1}), we obtain
	\begin{equation}
		\begin{split}
			\hat r\cdot\left\{4(\vec\rho\cdot\dot{\vec\rho})\vec\rho + (1-\rho^2)\left[D\dot{\vec\rho}+2(D-1)\beta\vec\rho -\frac{D}{2}(1+\rho^2)\textbf{K}\vec p -D\textbf{W}\vec\rho \right] \right\}\\ -2(\vec\rho\cdot\dot{\vec\rho})(1+\rho^2) +(1-\rho^2)[-D\vec\rho\cdot\dot{\vec\rho} + D\vec\rho\cdot\textbf{K}\vec p - (D-1)\beta(1+\rho^2)]=0\, ,
		\end{split}
	\end{equation}
	
	Here, the terms containing the angular coordinates, are written in the first line of the equation and the linear terms, in the second line. 
	Then, for this equation to be identically zero for each direction $\hat r$ this two parts must be independently zero.
	
	Considering the angular part, the terms inside the curly brackets leads us to
	\begin{equation}
		\dot{\vec\rho} = \textbf{W}\vec\rho + \frac{(1+\rho^2)}{2}\textbf{K}\vec p - \frac{2(D-1)}{D}\beta\vec\rho - \frac{4(\vec\rho\cdot\dot{\vec\rho})\vec\rho}{D(1-\rho^2)}
		\label{eqmotionvec}
	\end{equation}
	and $\dot{\vec\rho} =0$ when $\rho=1$.
	
	Noting that $\vec\rho\cdot\dot{\vec\rho} = \rho\dot\rho$ the linear part gives us 
	\begin{equation}
		[2+D+(2-D)\rho^2]\rho\dot\rho = (1-\rho^2)[D\vec\rho\cdot\textbf{K}\vec p - (D-1)\beta(1+\rho^2)].
		\label{eqmotionmod}
	\end{equation}
	
	Also, to ensure the compatibility of these two equations, we take the scalar product of Eq.(\ref{eqmotionvec}) with $\vec\rho$ and compare with Eq.(\ref{eqmotionmod}). Leaving all the calculations to Appendix \ref{app2} we find
	\begin{equation}
		\beta =\frac{(D-2)}{2(D-1)}\vec\rho\cdot\textbf{K}\vec p.
		\label{beta}
	\end{equation}
	
	Replacing $\beta$ on both Eq.(\ref{eqmotionvec}) and (\ref{eqmotionmod}), we find the equations of motion
	\begin{equation}
		\begin{split}
			\dot{\vec{\rho}} = \textbf{W} \vec{\rho}  + \frac{1}{2} (1+\rho^2) ({\mathbf K} \vec{p})  -  [\vec{\rho} \cdot ({\mathbf K} \vec{p}) ] \vec{\rho} .
		\end{split}
		\label{eqm1f}
	\end{equation}
	\begin{equation}
		\begin{split}
			\dot{\rho}  = \frac{1}{2}(1-\rho^2)  [\hat{\rho} \cdot ({\mathbf K} \vec{p}) ].
		\end{split}
		\label{eqm2f}
	\end{equation}  
	Notice that, remarkably, the dimensional dependence again cancels out and we have the same dynamics of the order parameter for any dimension. Replacing (\ref{beta}) into (\ref{coupling}) we see that the change in the vector $\vec{q}$ that ensures that the continuity equation is satisfied is
	\begin{equation}
		\vec q \rightarrow \vec{q} + \frac{D-2}{2(D-1)} [(\vec{\rho}-\hat{r}) \cdot \textbf{K}\vec p]  \hat r .
		\label{couplingtot}
	\end{equation}
	%
	
	
	\section{Equilibrium analysis}
	
	\subsection{Even dimensions}
	
	In order to look for equilibrium solutions, we consider the case where ${\mathbf K} = K\, \mathbb{1}$.  We start with the four-dimensional system. In this case, the unitary vector is given by $\vec x = (\sin\beta\sin\theta\cos\phi, \sin\beta\sin\theta\sin\phi, \sin\beta\cos\theta, \cos\beta)$ and the matrix of natural frequencies is 
	\begin{equation}
		\mathbf{W}_4 = \left( 
		\begin{array}{cccc}
			0 & -\omega_6 & \omega_5 & -\omega_4\\
			\omega_6 & 0 & -\omega_3 & \omega_2\\
			-\omega_5 & \omega_3 & 0 & -\omega_1\\
			\omega_4 & -\omega_2 & \omega_1 & 0\\
		\end{array}
		\right).
		\label{w4mat}
	\end{equation}
	
	If, without loss of generality, we choose the direction of the order parameter as $\vec p = p\hat x_4$ and write the ansatz vector as $\vec\rho = \alpha_1\hat x_1+ \alpha_2\hat x_2+ \alpha_3\hat x_3 +\alpha_4\hat x_4$, then the dynamics of $\rho$ is given by the Eq.(\ref{eqm1f}) and (\ref{eqm2f}), with $\vec\rho\cdot\vec p= p\,\alpha_4$.
	
	At equilibrium we have that 
	\begin{itemize}
		\item If $K=0$ then $\rho= p= 0$.
		\item If $K\neq 0$ and $\rho\neq 0$, then $\rho=1$ or $\vec{\rho}$ is perpendicular to $\vec{p}$, as we can see looking at Eq.(\ref{eqm2f}).
	\end{itemize}
	
	Let us first consider $\rho=1$. Then, Eq.(\ref{eqm1f}) becomes 
	\begin{equation}
		\begin{split}
			\dot{\vec{\rho}} = \textbf{W}_4 \vec{\rho}  +( K \vec{p})  -  [\vec{\rho} \cdot (K \vec{p}) ] \vec{\rho} 
		\end{split}
		\label{equilibvec}
	\end{equation}   
	or, in terms of components
	\begin{equation}
		\begin{split}
			\dot{\alpha}_1 &= -\omega_6\alpha_2 + \omega_5\alpha_3 -\omega_4\alpha_4 -Kp\alpha_4\alpha_1\\
			\dot{\alpha}_2 &= \omega_6\alpha_1 -\omega_3\gamma +\omega_2\delta -Kp\alpha_4\alpha_2\\
			\dot{\alpha}_3 &= -\omega_5\alpha_1 + \omega_3\alpha_2 -\omega_1\alpha_4 -Kp\alpha_4\alpha_3\\
			\dot{\alpha}_4 &= \omega_4\alpha_1 -\omega_2\alpha_2 +\omega_1\alpha_3 +Kp(1-\alpha_4^2).
		\end{split}
	\end{equation}   
	
	Setting  $\dot\alpha_1=\dot\alpha_2=\dot\alpha_3=\dot\alpha_4=0$ we find 
	\begin{equation}
		\alpha_4 = \sqrt{\frac{1}{2}-\frac{\omega^2 }{2 K^2 p^2} +   \sqrt{\left(\frac{1}{2}-\frac{\omega^2 }{2 K^2 p^2}\right)^2 + \frac{\xi}{K^2 p^2}}},
		\label{delta}
	\end{equation}
	where
	\begin{equation}
		\xi = K^2 p^2 (\omega_3^2 + \omega_5^2 + \omega_6^2) - (\omega_3 \omega_4 - \omega_2 \omega_5 + \omega_1 \omega_6)^2
	\end{equation}
	and
	\begin{equation}
		\omega^2 = \sum_{k=1}^6 \omega_k^2.
	\end{equation}
	
	Expressions for $\alpha_1$, $\alpha_2$ and $\alpha_3$ are more complicated and we shall not write them down. According to Eq.(\ref{paramFinal})
	\begin{equation}
		p = \int  \sqrt{\frac{1}{2}-\frac{\omega^2 }{2 K^2 p^2} +   \sqrt{\left(\frac{1}{2}-\frac{\omega^2 }{2 K^2 p^2}\right)^2 + \frac{\xi}{K^2 p^2}}   } \; G(\vec{\omega})\, d^6 \omega,
		\label{p4deq}
	\end{equation}
	where the integration is restricted to the region where $\alpha_4$ is real.
	
	Two important results can be derived from expression: first, for identical oscillators with $\vec{\omega}=0$, $p=1$. This is a trivial result implying the full synchronization of  the identical oscillators for $K>0$. Second, and more interesting, is the case where  $\omega_1=\omega_2=\omega_4=0$ and $\omega_3$, $\omega_5$ and $\omega_6$ are distributed according to $G$. In this case it is easy to check that $\alpha_4=1$ and, therefore, $p=1$, implying instantaneous synchronization of non-identical oscillators. This solution exists in all even dimensions (except $D=2$) and we show the analysis in Appendix \ref{app4}.
	\begin{figure*}
		\centering 
		\includegraphics[scale=0.5]{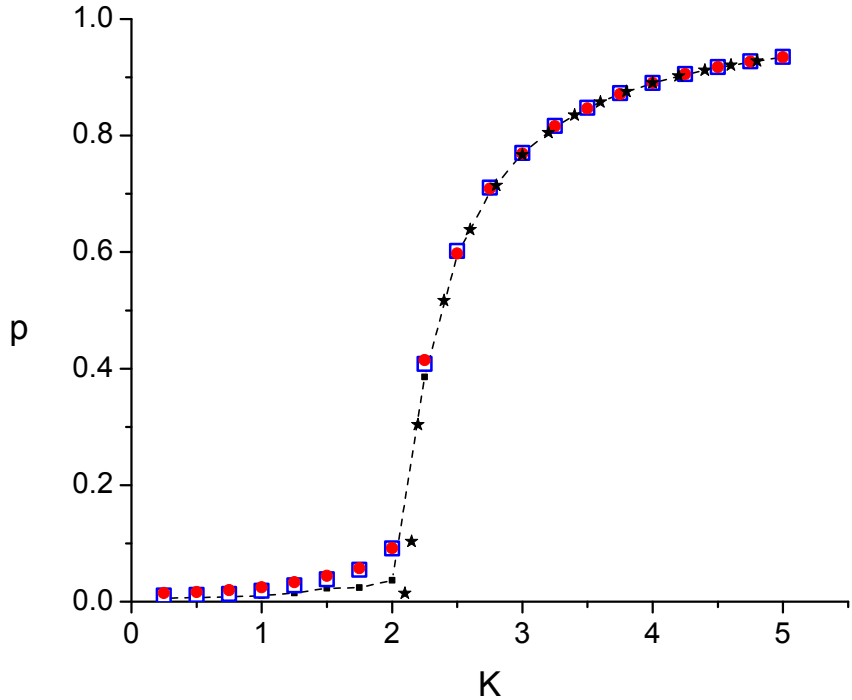} 
		\caption{Order parameter as a function of the coupling constant for a Gaussian distribution $G(\textbf{W})$ of natural frequencies and $D=4$: numerical integration of Eq. (\ref{3dkura}) with $N=30000$ oscillators (black squares and dashed line); numerical solution of Eq.(\ref{p4deq}) (stars); solution via our ansatz,  Eqs.(\ref{eqm1f}) and (\ref{paramFinal}) (red circles); solutions via the ansatz in \cite{chandra2019complexity} (blue squares). For both ansatz we discretized the integral relating $\vec{p}$ and $\vec{\rho}$ with $N=5000$ sets of frequencies, selected according to $G(\textbf{W})$.}  
		\label{fig1}
	\end{figure*}
	
	We complete our equilibrium analysis by finding the critical value $K_c$ for which the phase transition occurs for the case of Gaussian distribution of frequencies. Figure \ref{fig1} shows an example for $D=4$, comparing the numerical integration of the equations of motion for 30000 oscillators (red circles) and the result provided by Eq.(\ref{p4deq}) for a Gaussian distribution of each natural frequencies $\omega_k$, centered in zero with unit variance. 
	
	Based on the treatment given in \cite{chandra2019continuous}, we know that, for every even dimensional anti-symmetric matrix $\mathbf{W}$, there exists a real orthogonal matrix $\mathbf{R}$, such that  $\mathbf{R}^T \mathbf{WR}$ is a block-diagonal matrix whose $j$th block is the $2\times2$ matrix 
	\begin{equation}
		\mathbf{W}^{j} = \left( 
		\begin{array}{cc}
			0 & \tilde\omega_j\\
			-\tilde\omega_j & 0\\
		\end{array}
		\right),
		\label{blockmat}
	\end{equation}
	where $j \in \{1,\hdots, d\}$ with $d=D/2$. Then, for each dimension D we associate $D/2$ independent frequencies. Let $\{e_i\}_{i=1,\hdots,D}$ be the basis where $\mathbf{W}$ is block diagonal. Then, choosing $\vec p= p\hat e_D$ and $\vec\rho=\sum_i\alpha_i\hat e_i $, the components of the Eq.(\ref{eqm1f}) in this basis are
	\begin{equation}
		\begin{split}
			0&=\dot{\alpha}_1 = -\tilde\omega_1\alpha_2 - Kp\alpha_D\alpha_1\\
			0&=\dot{\alpha}_2 = \tilde\omega_1\alpha_1 - Kp\alpha_D\alpha_2\\
			\vdots\\
			0&=\dot{\alpha}_{D-3} = -\tilde\omega_{d-1}\alpha_{D-2} - Kp\alpha_D\alpha_{D-3}\\
			0&=\dot{\alpha}_{D-2} = \tilde\omega_{d-1}\alpha_{D-3} - Kp\alpha_D\alpha_{D-2}\\
			0&=\dot{\alpha}_{D-1} = -\tilde\omega_{d}\alpha_{D} - Kp\alpha_D\alpha_{D-1}\\
			0&=\dot{\alpha}_{D} = \tilde\omega_{d}\alpha_{D-1} + Kp(1- \alpha_D^2).
		\end{split}
	\end{equation}   
	Using $\rho=1$ the solution is
	\begin{equation}
		\begin{split}
			\alpha&_1=\hdots=\alpha_{D-2}=0\\
			\alpha&_{D-1}=\frac{\tilde\omega_{d}}{Kp}\\
			\alpha&_D=\left(1-\frac{\tilde\omega_d^2}{(Kp)^2}\right)^{\frac{1}{2}}.
		\end{split}
	\end{equation}  
	
	Then, 
	\begin{equation}
		p=\int \alpha_D \, G(\mathbf{W}) d \mathbf{W} =\int \left(1-\frac{\tilde\omega_{d}^2}{(Kp)^2}\right)^{\frac{1}{2}}g(\tilde\omega_1, \hdots, \tilde\omega_d)U(\mathbf{R}) d \tilde{\omega} d\mathbf{R}.
		\label{orderparameq}
	\end{equation} 
	The integral over $\tilde{\omega}_d$ is restricted to the interval $(-Kp,Kp)$ where $\rho=1$. Changing $\tilde{\omega}_d =Kp \,\xi$ we obtain 
	\begin{equation}
		p=p K_c  \int_{-1}^{1} d\xi  (1-\xi^2)^{1/2} \int g(\tilde\omega_1, \hdots, \tilde\omega_{d-1},\xi K p)U(\mathbf{R})  d \tilde{\omega}_1 d \tilde{\omega}_{d-1} d\mathbf{R}.
		\label{orderparameq1}
	\end{equation} 
	
	Canceling a factor of $p$ on each side and setting $p=0$ to find the critical point we obtain
	\begin{equation}
		1= K_c \int_{-1}^{1} d\xi (1-\xi^2)^{1/2} \int g(\tilde\omega_1, \hdots, \tilde\omega_{d-1},0) d \tilde{\omega}_1 d \tilde{\omega}_{d-1} \int U(\mathbf{R}) d\mathbf{R}
		\label{orderparameq2}
	\end{equation} 
	or 
	\begin{equation}
		1= K_c  \frac{\pi}{2}  \tilde{g}(0),
		\label{orderparameq3}
	\end{equation} 
	where 
	\begin{equation}
		\tilde{g}(x)=\int_0^\infty\hdots\int_0^\infty g(x_1, \hdots, x_d)dx_1\hdots dx_d 
		\label{}
	\end{equation} 
	and \cite{MEHTA1968449} 
	\begin{equation}
		\tilde{g}(0)=\frac{1}{D}\sqrt{\frac{2}{\pi}}\sum_{n=0}^{d-1}\frac{(2n)!}{2^{2n}(n!)^2}.
		\label{}
	\end{equation} 
	Therefore, as obtained by \cite{chandra2019continuous} we find
	\begin{equation}
		K_c=\frac{2}{\pi\tilde g(0)}.
	\end{equation}

	
	\subsection{Odd dimensions}
	As in the even dimensions equilibrium analysis, we consider ${\mathbf K} = K {\mathbf 1}$ and the equations of motion are given by Eq.(\ref{eqm1f}) and (\ref{eqm2f}). 
	
	Notice that the determinant of a $D\times D$ skew-symmetric matrix satisfies 
	\begin{equation}
		det(A) = det(A^\top)= det(-A) = (-1)^D det(A).
		\label{det}
	\end{equation} 
	
	So if $D$ is odd, the determinant vanishes. This means that for odd dimensions we always have an eigenvector with null eigenvalue. We will denote this eigenvector by $\hat\omega$
	
	In order to find equilibrium solutions we first consider the limit $K\rightarrow 0^+$. For $p\neq 0$, equilibrium requires that $\rho=1$ and $\hat\rho$ be parallel to $\hat\omega$, i.e., $\vec\rho_{eq}= \pm\hat\omega$. Now we must find out which of these solutions is the stable one. 
	
	First we add a perturbation to our equilibrium state $\vec\rho= \hat\rho_{eq} + \vec\epsilon(t)$, or $\rho= 1 + \epsilon(t)$. Then, Eq.(\ref{eqm2f}) to the first order gives us that the stable solution is the one for which $\hat\rho\cdot\vec p> 0$, as it is shown in Appendix \ref{app4},  Eq.(\ref{stability}). This means that $\hat{\rho}$ is in the hemisphere that is defined by $\vec p$. 
	
	Setting $\vec p= p\hat x_D$, we have $\rho_D = \hat\omega \cdot \hat x_D= \cos\theta_{D-1}$ and from Eq.(\ref{paramFinal}) the module of the order parameter is given by
	\begin{equation}
		p=\int\frac{G\mathbf{W}}{\mathcal{A}_D}\,\rho_D\, d\mathbf{W} = \frac{\mathcal{A}_{D-1}}{\mathcal{A}_D}\int_0^{\frac{\pi}{2}}2\sin^{D-2}\theta_{D-1}\cos\theta_{D-1}d\theta_{D-1},
		\label{p_eq}
	\end{equation} 
	
	where the factor 2 comes from the fact that, in the upper hemisphere, the stable solution is $\vec\rho=\hat\omega$ and in the lower hemisphere it is $\vec\rho= -\hat\omega$. So when we cross from one hemisphere to another both $\cos\theta_{D-1}$ and $\vec\rho$ changes signal, making the integral over $0 < \theta < \pi$ symmetrical
	with respect to $\frac{\pi}{2}$.
	
	Remember that 
	\begin{equation}
		\frac{\mathcal{A}_{D-1}}{\mathcal{A}_D} = \frac{2^{D-2}[\Gamma(\frac{D}{2})]^2}{\Gamma(D-1)\pi}.
	\end{equation} 
	
	And we can easily see, performing an integration by parts, that
	\begin{equation}
		\int_0^{\frac{\pi}{2}} 2\sin^{D-2}\theta_{D-1}\cos\theta_{D-1}d\theta_{D-1} = \frac{2}{D-1}.
	\end{equation} 
	
	Then, when $K\rightarrow 0^+$
	\begin{equation}
		p= \frac{2^{D-1}[\Gamma(\frac{D}{2})]^2}{(D-1)!\pi}= \frac{2\Gamma(\frac{D}{2})}{(D-1)\sqrt{\pi}\Gamma(\frac{D-1}{2})}, 
	\end{equation} 
	where we used the Legendre duplication formula for the last step. This is exactly what was obtained in \cite{chandra2019continuous}.
	
	
	\section{Numerical simulations}
	
	As an example of dynamic behavior we consider the case $D=4$ with Gaussian distribution of all six natural frequencies, centered around zero with unit variance (see Fig. \ref{fig1}). 
	
\begin{figure*}
		\centering 
		\includegraphics[scale=0.25]{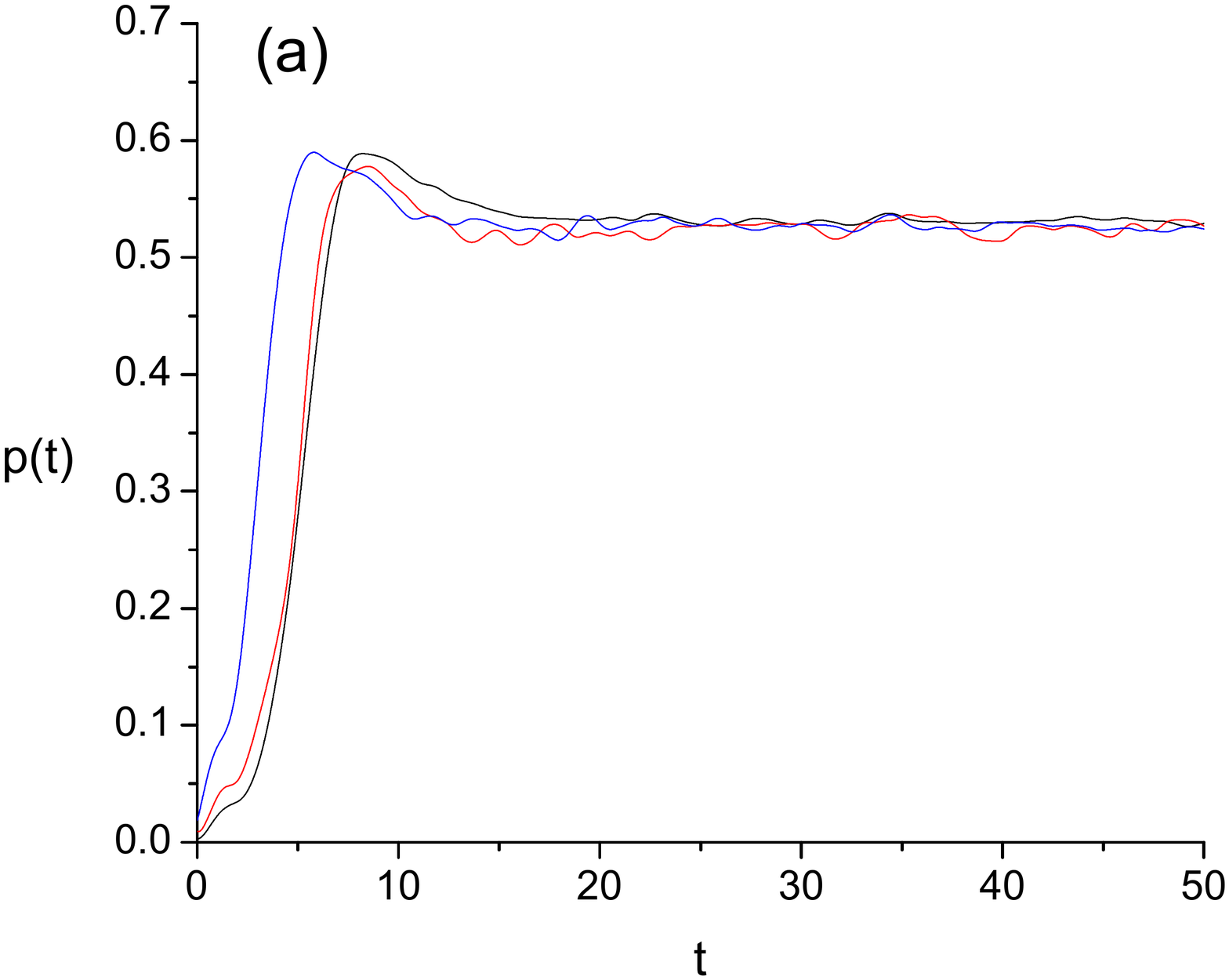} 
		\includegraphics[scale=0.25]{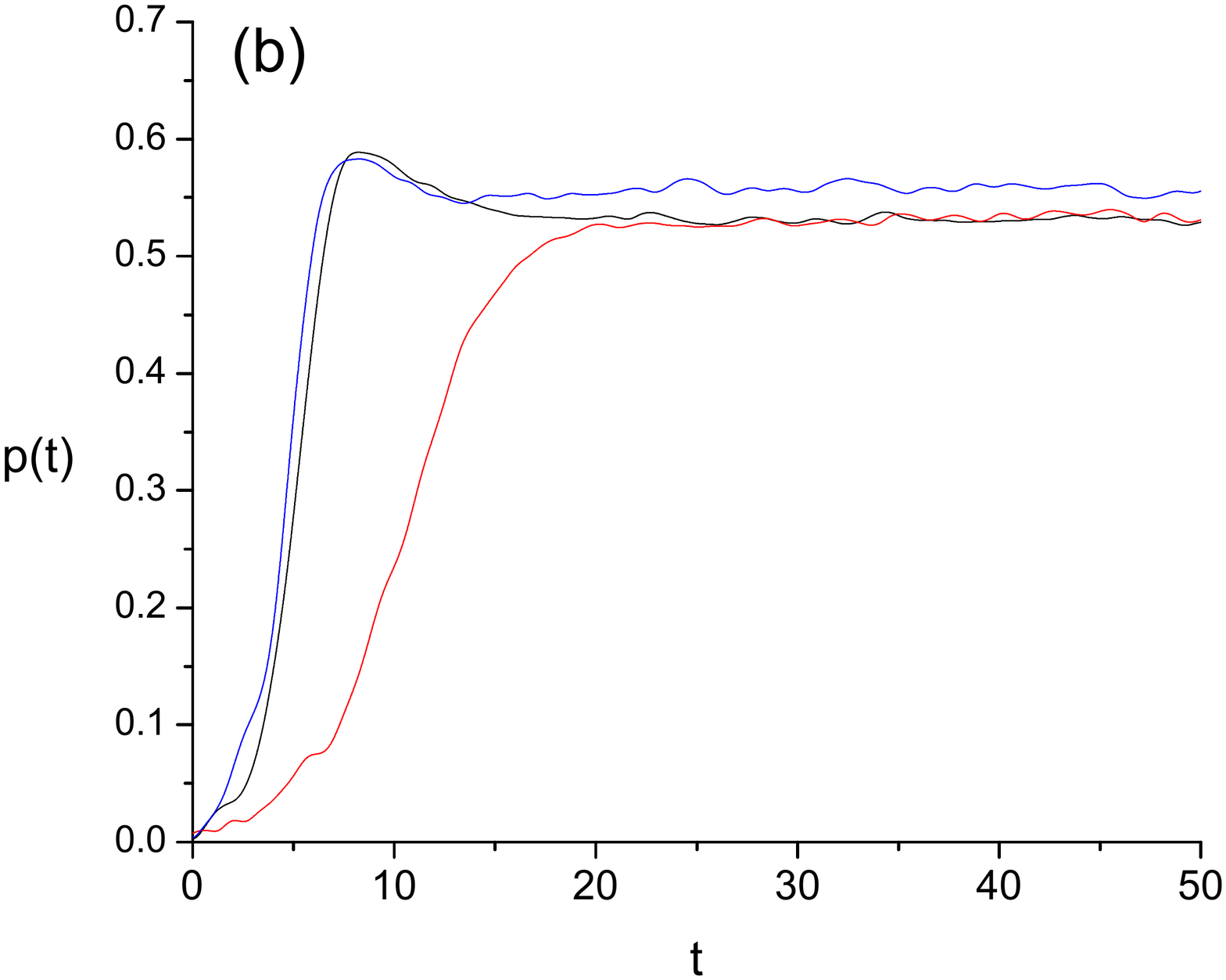} 
		\caption{Order parameter as a function of time for $K=2.4$ and Gaussian distribution of natural frequencies; (a) initial conditions satisfying $\rho_i(\textbf{W},0)=1$ for both PA and CGO (b) same simulations with initial conditions chosen randomly in the interval $0<\rho_i(\textbf{W},0)<1$. The black line shows the result of simulations with $N=30000$ oscillators whereas the red and blue lines shows the numerical solution using the PA and CGO ansatz respectively.}
		\label{fig2}
	\end{figure*}

	Figure \ref{fig2}(a) shows a comparison between the exact numerical calculation with $N=30,000$ oscillators (black line), the present ansatz (PA, for short, red line) and Chandra's et all proposal \cite{chandra2019complexity} (CGO, for short, blue line). The red and blue curves were computed using Eq.(\ref{eqm1f}) together with
	\begin{equation}
		\vec p(t) = \int G(\textbf{W})\,\vec\rho\ (\textbf{W},t) \, d\textbf{W}.
		\label{paramFinal2}
	\end{equation}
	for the PA and
	\begin{equation}
		\vec p(t) = \int G(\textbf{W})\, \left(\frac{3-\rho^2}{2} \right)\vec\rho\ (\textbf{W},t) \, d\textbf{W}.
		\label{paramFinalott}
	\end{equation}
	for CGO. In both cases the field $\vec\rho(\textbf{W},t)$ was discretized as $\vec\rho_i(t) = \vec\rho(\textbf{W}_i,t)$ where the $\textbf{W}_i$ were drawn according to $G(\textbf{W})$. We used $M=5000$ values of $\textbf{W}_i$ in our calculations and $\rho_i(\textbf{W},0)=1$ with random directions on the 4D sphere. We see that the approximate dynamics given by the ansatz equations follow very closely the numerical simulation in both cases. We note that the choice of unit initial module $\rho_i(\textbf{W},0)=1$ is not necessary. However, if the values of $\rho_i(\textbf{W},0)$ are chosen randomly in the interval $[0,1]$ the results are worse for PA, as shown if Fig. \ref{fig2}(b).
	
\begin{figure*}
		\centering 
		\includegraphics[scale=0.25]{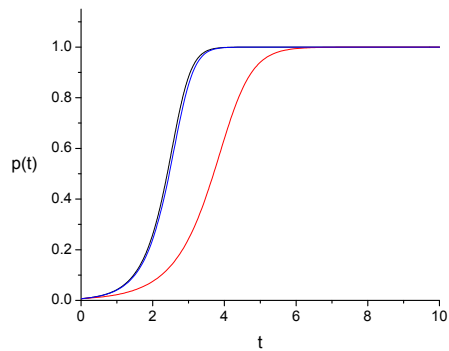} 
		\caption{Order parameter as a function of time for $K=2.4$ and identical oscillators with all natural frequencies set to $1$. The black line shows the result of simulations with $N=30000$ oscillators whereas the red and blue lines shows the numerical solution using the PA and CGO ansatz respectively.}
		\label{fig3}
	\end{figure*}

	Figure \ref{fig3} shows a similar comparison for identical oscillators, with $\textbf{W}$ given by Eq.(\ref{w4mat}) with all $\omega_i=1$. Here the equations simplify to a single differential equation and we obtain
	\begin{equation}
		\begin{split}
			\dot{\vec{p}} = \textbf{W} \vec{p}  +  \frac{K}{2} (1-p^2) \vec{p} 
		\end{split}
		\label{eqm1fid}
	\end{equation}
	for the PA and 
	\begin{equation}
		\begin{split}
			\dot{\vec{\rho}} = \textbf{W} \vec{\rho}  + \frac{K}{4} (1-\rho^2)(3-\rho^2)  \vec{\rho} 
		\end{split}
		\label{eqm1fidott}
	\end{equation}
	with
	\begin{equation}
		\vec{p} = \left(\frac{3-\rho^2}{2} \right)\vec\rho\
	\end{equation}
	for the CGO. The initial condition can now be set to match that of the simulation: first the $N= 30,000$ oscillators of the numerical simulation are randomly distributed over the surface of the 4D sphere with unit vectors $\vec\sigma_i$. The order parameter at time zero is then computed as $\vec p(0) = (1/N) \sum_i \vec \sigma_i$. For the CGO ansatz we set $\vec \rho (0) =2 \vec p(0) /3$. We see that, although both curves converge to $p=1$, the PA exhibits a delay that is corrected in the CGO ansatz by the extra kernel in (\ref{paramFinalott}). For other distributions $G(\textbf{W})$ this delay is compensated by the choice $\rho_i(\textbf{W},0)=1$.
	
\begin{figure*}
		\centering 
		\includegraphics[scale=0.19]{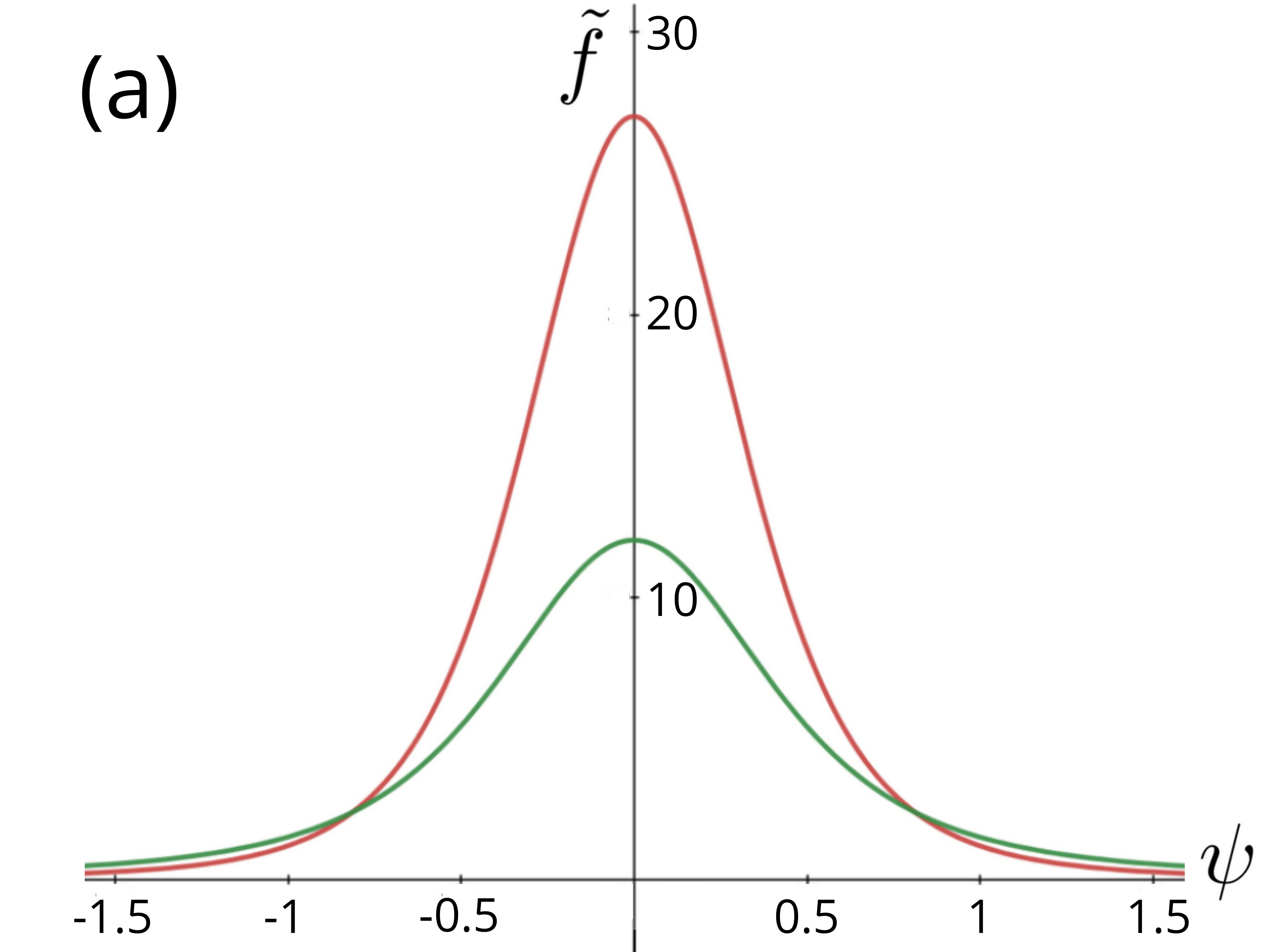} 
		\includegraphics[scale=0.19]{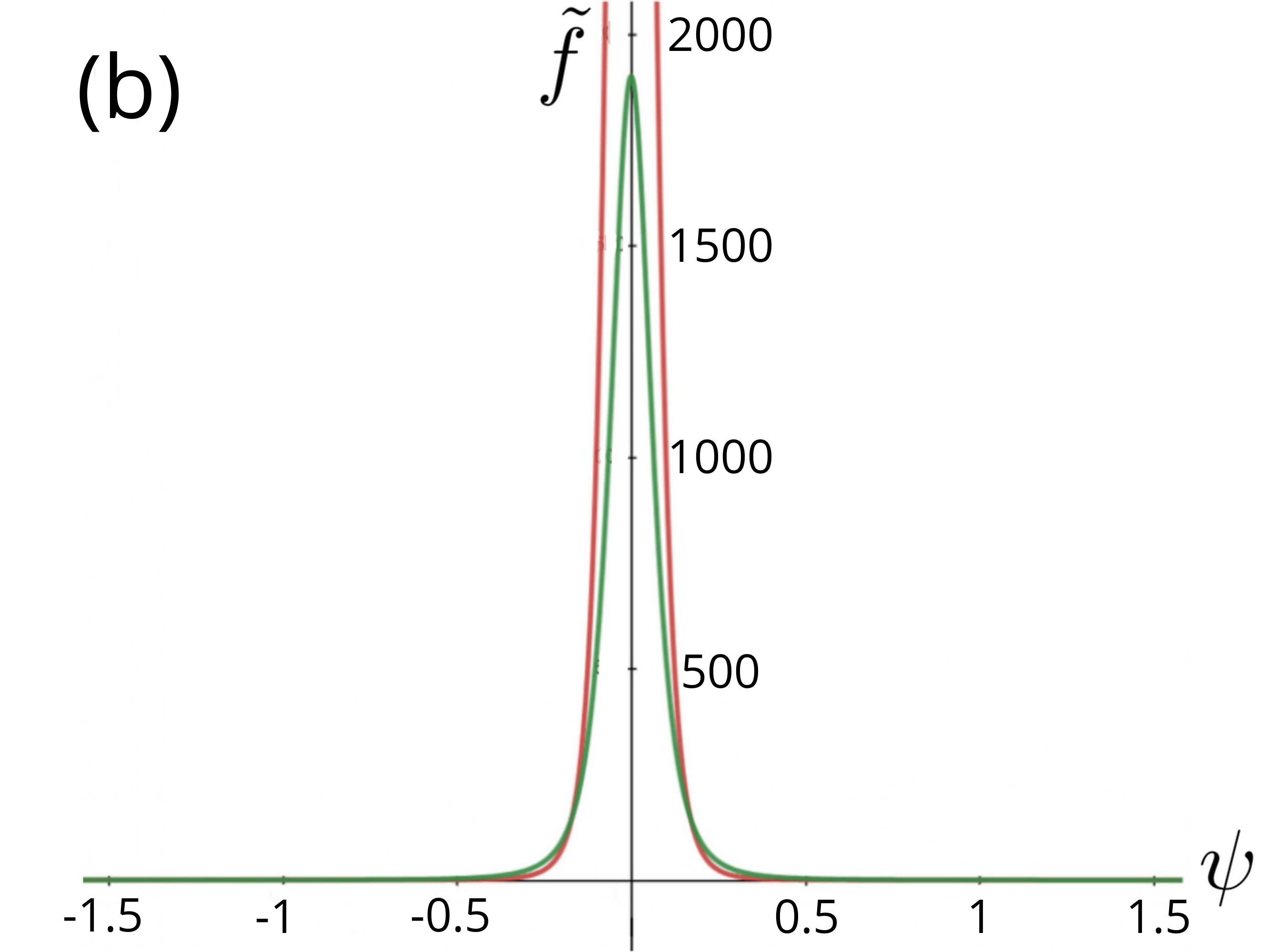} 
		\includegraphics[scale=0.19]{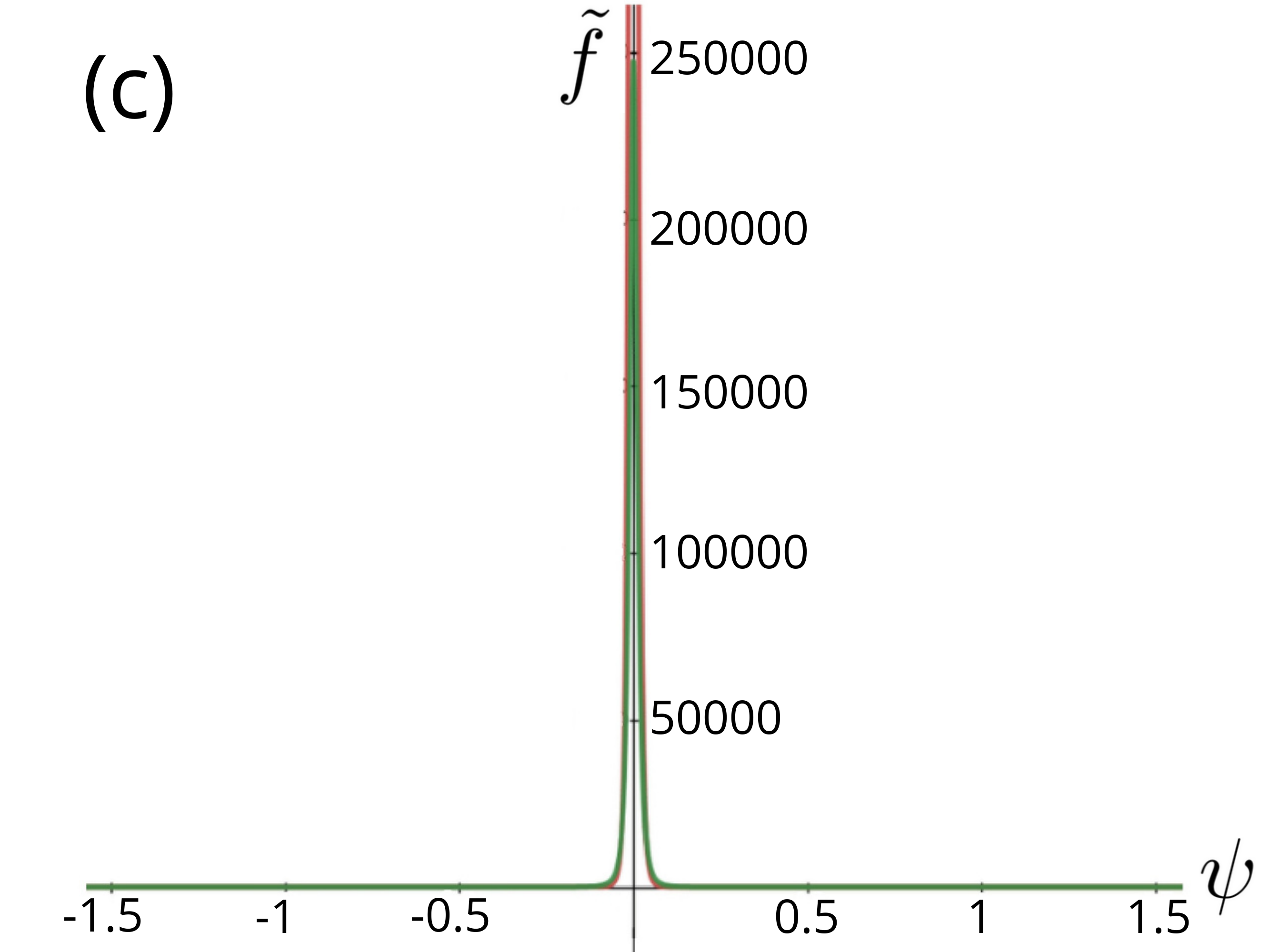} 
		\caption{Density distribution as a function of $\psi$; (a) for $\rho=0.5$; (b) for $\rho=0.9$; (c) for $\rho=0.98$. The red line is the distribution function as proposed by CGO and the green line represents the PA distribution.}
		\label{fig4}
	\end{figure*}

	The delay in the PA with respect to the CGO ansatz is a consequence of the difference between the density functions. In order to see that, we plotted in Fig. \ref{fig4}, for $D=4$,  both density functions in terms of $\hat\rho\cdot\hat r\equiv \cos{\psi}$. The PA distribution is given by Eq. (\ref{distransatz}) as
	\begin{equation}
		f_{PA}(\vec r, \textbf{W}, t) =G(\textbf{W})\frac{1}{2\pi^{2}} \frac{1-\rho^2}{(1+\rho^2-2\rho \cos\psi)^{2}} \equiv \frac{G(\textbf{W})}{2\pi^{2}} \tilde{f}_{PA}(\rho,\psi)
	\end{equation}
	and CGO density function by 
	\begin{equation}
		f(\vec r, \textbf{W}, t)_{CGO} = G(\textbf{W}) \frac{1}{2\pi^{2}} \frac{(1-\rho)^3}{(1+\rho^2-2\rho \cos\psi)^3}
		\equiv \frac{G(\textbf{W})}{2\pi^{2}} \tilde{f}_{CGO}(\rho,\psi).
	\end{equation}
	
	Figure shows \ref{fig4} that for small $\rho$, $\tilde{f}_{CGO}$ has a higher and sharper peak around $\psi=0$ than the $\tilde{f}_{PA}$, i.e., the equilibrium is approached more rapidly. But as $\rho$ increases, for example $\rho=0.9$ and $\rho=0.98$ it is the PA distribution that provides a sharper peak. This suggests that although the PA is delayed in comparison to CGO ansatz, it reaches the equilibrium faster after the transient delay. It is important to note that at the equilibrium, when $\rho=1$, the two methods agree and should provide a good approximation of the order parameter. Also, the initial condition $\rho=1$ for non-identical oscillators removes the delay in PA dynamics.

	\section{Conclusion}
	
	We have developed an alternative and equally accurate ansatz to solve the Kuramoto model in any dimension. Our approach is different from previously proposed formulations and consists in proposing an ansatz based on generalized spherical harmonics decomposition of the density function of the oscillator’s positions. The continuity equation for the ansatz generates undesired terms in the radial directions, which are eliminated by modifying the coupling term in that direction. Because such modification does not alter the original equations (as these terms are always canceled out automatically) the equations for the ansatz parameters are still approximations for the Kuramoto system.

	The dynamics we obtained from the continuity equation leads to a vector equation for the ansatz parameters that is exactly that obtained by Chandra el al's  ansatz \cite{chandra2019complexity}. However, the connection between the ansatz and the order parameter differs and it is simpler in our approach. The relationship we find is the natural extension of the 2 and 3 dimensional cases, allowing the interpretation of $\vec\rho(\vec\omega)$ as the order parameter for subset of oscillators with frequencies $\vec{\omega} = \{\omega_1, \hdots, \omega_{\frac{D(D-1)}{2}}\}$ and $\vec p$ as the average of $\vec\rho(\vec\omega)$ over the frequency distribution $G(\mathbf{W})$. This might facilitate the analysis of more general systems where the coupling is a full matrix $\mathbf{K}$ or in the presence of external forces. The complete dimensional reduction is only achieved for the case of identical oscillators. For all other cases, the connection between the ansatz and the order parameter can be solved by Monte Carlo methods, sampling the distribution $G(\mathbf{W})$, which converges fast for most distributions of interest. 
	
	We also analyzed the equilibrium conditions and the phase transitions on both even and odd dimensions, noticing that, as it have already been proved, for even dimensions the transition to coherence occurs continuously while for odd dimensions the transition is discontinuous. With that, we have shown that the critical value of the coupling constant for even D and the value of $p$ in the transition for odd D agree with previous calculations using the exact equations \cite{chandra2019continuous}. In addition to these cases, there is a particular choice of frequency that leads the system to instantaneous synchronization of non-identical oscillators. For $D=4$ we showed that it is possible to obtain semi-analytical solution for the critical curve $p=p(K)$, as shown by Eq.(\ref{p4deq}) and Fig. \ref{fig1}.
	

	
	\begin{acknowledgments}
		This work was partly supported by FAPESP, grants 2019/20271-5 (MAMA), 2019/24068-0 (AEDB), 2016/01343‐7 (ICTP‐SAIFR FAPESP) and CNPq, grant 301082/2019‐7 (MAMA). We would like to thank Alberto Saa and Jose A. Brum for suggestions and careful reading of this manuscript.
	\end{acknowledgments}
	

	\clearpage 
	\newpage
	
	\appendix
	
	\section{Generalized Spherical Harmonics and their properties}
	\label{app1}
	
	Based on what is already known about the spherical harmonics and on four-dimensional hyperspherical harmonics \cite{Domokos1967}, we propose a generalization to the arbitrary dimension, D. Here we summarize the principal definitions and properties of the Generalized Spherical Harmonics, but more details are available in ref.\cite{Avery2018}.

	We first notice that the Fourier series (equivalent of the spherical harmonics for 2D), the usual spherical harmonics in 3D and the 4D harmonics can be written in terms of the corresponding harmonic in the dimension immediately below as
	\begin{equation}
		\begin{split}
			Y^m_{l_2, \ldots, l_{D-1}}(\theta_1, \ldots, \theta_{l_{D-1}})=N(l_1, \ldots, l_{D-1})\sin^{l_{D-2}}\theta_{D-1}C^\alpha_{l_{D-1}-l_{D-2}}(\cos{\theta_{D-1}})\\
			\times Y^m_{l_2,\ldots, l_{D-2}}(\theta_1,\ldots, \theta_{D-2}), 
		\end{split}
	\end{equation}
	where $C^\alpha_n$ are the Gegenbauer polynomials and we defined $m\equiv l_1$ because this is the index associated with the only angle, $\theta_1$, that ranges over $[0,2\pi)$.
	
	It still remains to be determined the normalization constant $N(l_1, \ldots, l_{D-1})$ and the Gegenbauer index $\alpha$.
	In order to find the normalization, we integrate over the angles, remembering that the surface element is given by Eq.(\ref{VolEl})
	\begin{equation}
		\int_0^\pi\ldots\int_0^\pi\int_0^{2\pi} Y^m_{l_2, \ldots, l_{D-1}}Y^{m\,*}_{l'_2, \ldots, l'_{D-1}}d^DS \,=\, \delta_{l_1, l'_1}\ldots\delta_{l_{D-1}, l'_{D-1}}.
		\label{NormInt}
	\end{equation}
	
	When not all of the index are equal, we assume, by induction, the orthogonality of the previous dimension harmonic. For $l_i=l'_i$, $i\in\{1,\ldots, D-1\}$ the integral in Eq.(\ref{NormInt}) becomes
	\begin{equation}
		\begin{split}
			\int_0^\pi\ldots\int_0^\pi\int_0^{2\pi}N^2\sin^{2l_{D-2}}\theta_{D-1}[C^\alpha_{l_{D-1}-l_{D-2}}(\cos{\theta_{D-1}})]^2 Y^m_{l_2, \ldots, l_{D-2}}Y^{m\,*}_{l_2, \ldots, l_{D-2}}d^DS \,=\, 1 \\
			=\int_0^\pi N^2\sin^{2l_{D-2}}\theta_{D-1}[C^\alpha_{l_{D-1}-l_{D-2}}(\cos{\theta_{D-1}})]^2\sin^{D-2}\theta_{D-1}d\theta_{D-1}\,=\, 1.
		\end{split}
	\end{equation}

	From the orthogonality properties of the Gegenbauer polynomials, we have 
	\begin{equation}
		\int_0^\pi C_n^\alpha(\cos{\theta})C_{n'}^\alpha(\cos{\theta})\sin^{2\alpha}\theta d\theta =\frac{\pi 2^{1-2\alpha}\Gamma(n+2\alpha)}{n!(n+\alpha)[\Gamma(\alpha)]^2}\delta_{n,n'}.
		\label{GegenOrt}
	\end{equation}
	
	Therefore, in order to apply this relation, we establish the value of $\alpha$
	\begin{equation}
		\alpha = l_{D-2}-1+\frac{D}{2}.
	\end{equation}
	
	And the normalization constant becomes 
	\begin{equation}
		N^2_{l_{D-1}, l_{D-2}}=\frac{(l_{D-1}- l_{D-2})! (l_{D-1}-1+\frac{D}{2})}{\pi \Gamma(l_{D-1}- l_{D-2}+D-2)}2^{2l_{D-2}+D-3}\left[\Gamma(l_{D-2}-1+\frac{D}{2})\right]^2.
		\label{NormConst}
	\end{equation}
	
	Now we can write the final expression for the generalized spherical harmonic
	\begin{equation}
		Y^{m}_{l_2,\ldots,l_{D-1}}(\theta_1, \ldots, \theta_{D-1})= N_{l_{D-1},l_{D-2}}\sin^{l_{D-2}}{
			\theta_{D-1}}C_{l_{D-1}-l_{D-2}}^{l_{D-2}+\frac{D}{2}-1}(\cos{ \theta_{D-1}})Y^{m}_{l_2,\ldots,l_{D-2}}(\theta_1, \ldots, \theta_{D-2}),
		\label{HyperSH}
	\end{equation}
	
	\textbf{Lemma:} The hyperspherical harmonics with negative index $m$ are given by
	\begin{equation}
		Y^{-m}_{l_2,\ldots,l_{D-1}}= (-1)^m Y^{m\,*}_{l_2,\ldots,l_{D-1}}.
	\end{equation}
	
	\textbf{Proof:} Using induction we prove that
	
	(i) For D=3, we already have $Y^{-m}_l(\theta, \phi)=(-1)^m Y^{m\,*}_l(\theta, \phi)$
	
	(ii) If $Y^{-m}_{l_2,\ldots,l_{D-2}}= (-1)^m Y^{m\,*}_{l_2,\ldots,l_{D-2}}$, then for dimension D, we apply the definition of the hyperspherical harmonics, Eq.(\ref{HyperSH}) to obtain
	\begin{equation}
		\begin{split}
			Y^{-m}_{l_2,\ldots,l_{D-1}}&= N_{l_{D-1},l_{D-2}}\sin^{l_{D-2}}{\theta_{D-1}}C_{l_{D-1}-l_{D-2}}^{l_{D-2}+\frac{D}{2}-1}(\cos{ \theta_{D-1}})Y^{-m}_{l_2,\ldots,l_{D-2}}(\theta_1, \ldots, \theta_{D-2})\\
			&=N_{l_{D-1},l_{D-2}}\sin^{l_{D-2}}{
				\theta_{D-1}}C_{l_{D-1}-l_{D-2}}^{l_{D-2}+\frac{D}{2}-1}(\cos{ \theta_{D-1}})(-1)^m Y^{m\*}_{l_2,\ldots,l_{D-2}}(\theta_1, \ldots, \theta_{D-2})\\
			&= (-1)^m Y^{m\,*}_{l_2,\ldots,l_{D-1}}(\theta_1, \ldots, \theta_{D-1}).
		\end{split}
	\end{equation}

	\textbf{Lemma:} The D-hyperspherical harmonics satisfies
	\begin{equation}
		Y^0_{0,\ldots, 0}=\sqrt{\frac{\Gamma(\frac{D}{2})}{2\pi^{\frac{D}{2}}}}.
	\end{equation}
	
	\textbf{Proof:} Let's proceed by induction
	(i) For D=3, \quad $Y^0_0=\sqrt{\frac{1}{4\pi}} = \sqrt{\frac{\sqrt{\pi}}{2}\frac{1}{2\pi^{\frac{3}{2}}}}=\sqrt{\frac{\Gamma(\frac{3}{2})}{2\pi^{\frac{3}{2}}}}$
	(ii) If $Y^0_{0,\ldots, 0} =\sqrt{\frac{\Gamma(\frac{D}{2})}{2\pi^{\frac{D}{2}}}}$ for dimension D, then for dimension D+1
	\begin{equation}
		\underbrace{Y^0_{0,\ldots, 0}}_\text{dimension D+1} = \sqrt{\frac{D-1}{2\pi\Gamma(D-1)}}2^{\frac{D-2}{2}}\Gamma(\frac{D-1}{2})\underbrace{Y^0_{0,\ldots, 0}}_\text{dimension D}.
		\label{induction}
	\end{equation}
	
	Applying the induction hypothesis, 
	\begin{equation}
		\underbrace{Y^0_{0,\ldots, 0}}_\text{dimension D+1} = \sqrt{\frac{(D-1)\Gamma(\frac{D}{2})}{4\pi^{\frac{D}{2}+1}\Gamma(D-1)}}2^{\frac{D-2}{2}}\Gamma(\frac{D-1}{2}).
	\end{equation}
	
	Using the Legendre duplication formula, Eq.(\ref{induction}) becomes
	\begin{equation}
		\underbrace{Y^0_{0,\ldots, 0}}_\text{dimension D+1} = \sqrt{\frac{\Gamma(\frac{D+1}{2})2^{D-1}2^{2-D}}{4\pi^{\frac{D}{2}+\frac{1}{2}}}} = \sqrt{\frac{\Gamma(\frac{D+1}{2})}{2\pi^{\frac{D+1}{2}}}}.
		\label{Yzero}
	\end{equation}
	
	\vskip 1cm

	\textbf{Theorem (Generalized Addition Theorem):} Given two vectors $\hat r$ and $\hat \rho$, with coordinates $(r,\theta_1, \ldots, \theta_{D-1})$ and $(\rho,\theta'_1, \ldots, \theta'_{D-1})$, respectively, then the hyperspherical harmonics satisfies the relation
	\begin{equation}
		\sum_{l_{D-2}=0}^{l_{D-1}}\ldots\sum_{m=-l_2}^{l_2} Y^{m *}_{l_2,\ldots,l_{D-1}}(\hat\rho)Y^m_{l_2,\ldots,l_{D-1}}(\hat r) =\left(\frac{2l_{D-1}}{D-2}+1\right)\frac{C_{l_{D-1}}^\frac{D-2}{2}(\hat\rho\cdot\hat r)}{\mathcal{A}_D},
		\label{AddTheorem}
	\end{equation}
	where $\mathcal{A}_D$ is the area of the sphere $S_{D-1}$, which is given by
	\begin{equation}
		\mathcal{A}_D= \frac{2\pi^{\frac{D}{2}}}{\Gamma(\frac{D}{2})}.
		\label{AreaSphere}
	\end{equation}

	\textbf{Proof:} We can write the Gegenbauer polynomial $C^{\frac{D-2}{2}}_{l_{D-1}}$ as a series of hyperspherical harmonics
	\begin{equation}
		C^{\frac{D-2}{2}}_{l_{D-1}}(\hat\rho\cdot\hat r) = \sum_{l_{D-2}=0}^{l_{D-1}}\sum_{l_{D-3}=0}^{l_{D-2}}\ldots\sum_{m=-l_2}^{l_2} A^m_{l_2,\ldots,l_{D-1}}(\hat\rho)Y^m_{l_2,\ldots,l_{D-1}}(\hat r).
		\label{ExpansionC}
	\end{equation}
	
	Then, by the orthogonality of the hyperspherical harmonics, the coefficient of the expansion is given by
	\begin{equation}
		A^m_{l_2,\ldots,l_{D-1}}(\hat\rho) = \int Y^{m\,*}_{l_2,\ldots,l_{D-1}}(\hat r)C^{\frac{D-2}{2}}_{l_{D-1}}(\hat\rho\cdot\hat r)\, d^DS.
	\end{equation}
	
	We now choose the coordinates which represent the angles between $\hat r$ and $\hat\rho$. We denote this change of variables by $\theta_i\rightarrow\theta''_i$ and $d^DS\rightarrow d^DS''$. With that, the hyperspherical harmonics can be written as
	\begin{equation}
		Y^{m\,*}_{l_2,\ldots,l_{D-1}}(\hat r)\,=\,Y^{m\,*}_{l_2,\ldots,l_{D-1}}(\hat r(\theta''_1, \ldots, \theta''_{D-1}))\,=\, g(\theta''_1, \ldots, \theta''_{D-1}).
	\end{equation}
	
	We also expand $g(\theta''_1, \ldots, \theta''_{D-1})$ in a series of hyperspherical harmonics
	\begin{equation}
		g(\theta''_1, \ldots, \theta''_{D-1})= \sum_{l_{D-2}=0}^{l_{D-1}}\sum_{l_{D-3}=0}^{l_{D-2}}\ldots\sum_{m=-l_2}^{l_2} B^{m'}_{l'_2,\ldots,l'_{D-2}}Y^{m'}_{l'_2,\ldots,l'_{D-2}, l_{D-1}}(\theta''_1, \ldots, \theta''_{D-1}).
	\end{equation}
	
	The coefficient of the expansion is given by 
	\begin{equation}
		B^{m'}_{l'_2,\ldots,l'_{D-2}}\,=\,\int Y^{m'\,*}_{l'_2,\ldots,l'_{D-1}, l_{D-1}}(\theta''_1, \ldots, \theta''_{D-1})g(\theta''_1, \ldots, \theta''_{D-1})d^DS''.
	\end{equation}
	
	Notice that, if we want $B^{0}_{0,\ldots,0}$, we first calculate
	\begin{equation}
		Y^{0\,*}_{0,\ldots,0,l'_{D-1}}\,=\, \sqrt{\frac{l_{D-1}!(l_{D-1}+\frac{D}{2}-1)}{\pi\Gamma(l_{D-1}+D-2)}}2^{\frac{D-3}{2}}\Gamma(\frac{D-2}{2})C^{\frac{D-2}{2}}_{l_{D-1}}(\theta''_{D-1}) Y^{0*}_{0,\ldots,0}.
	\end{equation}
	
	Here we use the result of the proposition (\ref{Yzero}) for the $(D-1)$-hyperspherical harmonics we have
	\begin{equation}
		Y^{0\,*}_{0,\ldots,0,l'_{D-1}}\,=\, \sqrt{\frac{l_{D-1}!(l_{D-1}+\frac{D}{2}-1)2^{3-D}}{2\pi^{\frac{D-1}{2}}\Gamma(l_{D-1}+D-2)\Gamma(\frac{D-1}{2})}}\Gamma(D-2)
		C^{\frac{D-2}{2}}_{l_{D-1}}(\theta''_{D-1}).
	\end{equation}

	Now, we are able to write $B^{0}_{0,\ldots,0}$
	\begin{equation}
		\begin{split}
			B^{0}_{0,\ldots,0}\,&=\, \tilde{B}_{l_{D-1}} \int C^{\frac{D-2}{2}}_{l_{D-1}}(\theta''_{D-1})g(\theta''_1, \ldots, \theta''_{D-1})\\
			&=\, \tilde{B}_{l_{D-1}} A^m_{l_2, \ldots, l_{D-1}}, 
		\end{split}
	\end{equation}
	where we defined
	\begin{equation}
		\tilde{B}_{l_{D-1}}\sqrt{\frac{l_{D-1}!(l_{D-1}+\frac{D}{2}-1)2^{3-D}}{2\pi^{\frac{D-1}{2}}\Gamma(l_{D-1}+D-2)\Gamma(\frac{D-1}{2})}}\Gamma(D-2).
	\end{equation}

	So, for $\theta''_{D-1}=0$, we have
	\begin{equation}
		g(\theta''_1, \ldots, \theta''_{D-2}, 0)\,=\,Y^{m'\,*}_{l'_2,\ldots,l'_{D-2}, l_{D-1}}(\theta''_1, \ldots, \theta''_{D-2}, 0)\,=\, \tilde{B}_{l_{D-1}}C^{\frac{D-2}{2}}_{l_{D-1}}(1).
		\label{ThetaZero}
	\end{equation}
	
	We use the relation from \cite{SanKim2012} 
	\begin{equation}
		C^\alpha_n(1)\,=\, \frac{(n+\alpha-1)!}{n!(2\alpha-1)!}.
	\end{equation}
	
	Substituting this into Eq.(\ref{ThetaZero}), this leads us to
	\begin{equation}
		Y^{m'\,*}_{l'_2,\ldots,l'_{D-2},l_{D-1} }(\theta''_1, \ldots, \theta''_{D-2}, 0)\,=\, \tilde{B}_{l_{D-1}}\frac{(l_{D-1}+D-3)!}{l_{D-1}!(D-3)!}\prod_{i=1}^{D-2}\delta{l'_i, 0}.
	\end{equation}
	
	That way $g(\theta''_1, \ldots, \theta''_{D-2}, 0)$ can be written as
	\begin{equation}
		g(\theta''_1, \ldots, \theta''_{D-2}, 0)\,=\, \left(\frac{2l_{D-1}}{D-2}+1\right)\frac{A^m_{l_2, \ldots, l_{D-1}}}{\mathcal{A}_D}, 
	\end{equation}
	where we have used the Legendre duplication formula and $\mathcal{A}_D$ is again the area of the sphere $S_{D-1}$, which is given by Eq.(\ref{AreaSphere})
	
	Notice that if $\theta''_{D-1}=0$, then $\theta''_i=\theta'_i$ for all $i\in \{1,\ldots, l_{D-2}\}$. Now, 
	\begin{equation}
		Y^{m\,*}_{l_2,\ldots,l_{D-1}}(\hat\rho)\,=\, \left(\frac{2l_{D-1}}{D-2}+1\right)\frac{A^m_{l_2, \ldots, l_{D-1}}}{\mathcal{A}_D}.
	\end{equation}
	
	Finally, substituting this into Eq.(\ref{ExpansionC}) we have the theorem 
	\begin{equation}
		C^{\frac{D-2}{2}}_{l_{D-1}}(\hat\rho\cdot\hat r)\,=\, \frac{\mathcal{A}_D}{\left(\frac{2l_{D-1}}{D-2}+1\right)}\sum_{l_{D-2}=0}^{l_{D-1}}\ldots\sum_{m=-l_2}^{l_2} Y^{m *}_{l_2,\ldots,l_{D-1}}(\hat\rho)Y^m_{l_2,\ldots,l_{D-1}}(\hat r).
	\end{equation}

	\textbf{Proposition:} Let $\mathcal{C}_D=\{x_1, \ldots, x_D\}$ be the set of spherical coordinates in dimension D with $r=1$. Then all the terms of $\mathcal{C}_D$ can be written in terms of the hyperspherical harmonics with $l_{D-1}=1$, $Y^m_{l_2,\ldots, l_{D-2}, 1}$.
	
	\textbf{Proof:} First, let $l_2,\ldots, l_{D-2}=0$. Then
	\begin{equation}
		Y^m_{0,\ldots, 0, 1}=\sqrt{\frac{D2^{1-D}}{\pi^{\frac{D-1}{2}}\Gamma(D-1)\Gamma(\frac{D-1}{2})}}2(D-1)!\cos{\theta_{D-1}}.
	\end{equation}
	
	Notice that $x_D=\cos{\theta_{D-1}}$ can then be written in terms of $ Y^m_{0,\ldots, 0, 1}$. Now, we prove, by induction, that the other coordinates can be written in terms of $Y^m_{l_2,\ldots, l_{D-2}, 1}$. First, notice that for $D=3$ this is true because 
	\begin{equation}
		\begin{split}
			x_1 &= \sin{\theta}\cos{\phi} = \sqrt{\frac{2\pi}{3}}(Y^{-1}_1-Y^{1}_1)\\
			x_2 &= \sin{\theta}\sin{\phi}=i\sqrt{\frac{2\pi}{3}}(Y^{-1}_1+Y^{1}_1)\\
			x_3 &= \cos{\theta} = -2\sqrt{\frac{\pi}{3}}Y^0_1.
		\end{split}
	\end{equation}
	
	Now, we suppose that, for dimension $D-1$, we can write all the elements of $\mathcal{C}_{D-1}$ in terms of $Y^m_{l_2,\ldots, l_{D-3}, 1}$. Then, 
	notice that
	\begin{equation}
		Y^m_{0,\ldots, l_{D-3}, 1, 1}=\sqrt{\frac{D}{2\pi\Gamma(D)}}2^{\frac{D-1}{2}}\Gamma\left(\frac{D}{2}\right)\sin{\theta_{D-1}}Y^m_{0,\ldots, l_{D-3}, 1}.
	\end{equation}
	
	Since the coordinates $x_1, \ldots, x_{D-1}$ are given by the elements of $\mathcal{C}_{D-1}$ multiplied by $\sin{\theta_{D-1}}$, by the induction hypothesis, we conclude that all the elements of $\mathcal{C}_D$ can be written in terms of the hyperspherical harmonics with $l_{D-1}=1$, $Y^m_{l_2,\ldots, l_{D-2}, 1}$.
	
	
	
	\section{Connection between angular and linear parts of the continuity equation}
	\label{app2}
	
	In order to find the appropriate $\beta$ that makes Eq.(\ref{eqmotionvec}) and Eq.(\ref{eqmotionmod}) consistent, first notice that $\vec\rho\cdot\textbf{W}\vec\rho = 0$, as we can see by 
	\begin{equation}
		\begin{split}
			\vec\rho^{\,\,T}\textbf{W}\vec\rho\, =\, \vec\rho\cdot\textbf{W}\vec\rho\, =\, \textbf{W}\vec\rho\cdot\vec\rho\, =\,  \vec\rho^{\,\,T}\textbf{W}^{\,T}\vec\rho\, =\, \vec\rho^{\,\,T}(-\textbf{W})\vec\rho\, =\, -\vec\rho^{\,\,T}(\textbf{W})\vec\rho.
		\end{split}
	\end{equation}
	
	Then, we take the scalar product of Eq.(\ref{eqmotionvec}) with $\vec\rho$ 
	\begin{equation}
		\left[\frac{D + (4-D)\rho^2}{D(1-\rho^2)}\right]\rho\dot\rho = \frac{1}{2} (1+\rho^2) \vec\rho\cdot{\mathbf K} \vec{p} - \frac{2(D-1)}{D}\beta\rho^2.
	\end{equation}
	Now we substitute $\rho\dot\rho$ from Eq.(\ref{eqmotionmod})
	\begin{equation}
		\begin{split}
			\left[\frac{D+(4-D)\rho^2}{D(1-\rho^2)}\right]&\left[\frac{(1-\rho^2)}{2+D+(2-D)\rho^2}\right](D\vec\rho\cdot\textbf{K}\vec p - (D-1)(1+\rho^2)\beta) \\
			&= \frac{1}{2} (1+\rho^2) \vec\rho\cdot{\mathbf K} \vec{p} - \frac{2(D-1)}{D}\beta\rho^2.
		\end{split}
	\end{equation}
	
	Simplifying this equation, we obtain
	\begin{equation}
		\frac{(D-2)(1-2\rho^2+\rho^4)}{2(2+D+(2-D)\rho^2)}\, \vec\rho\cdot\textbf{K}\vec p = \frac{(D-1)(1-2\rho^2+\rho^4)}{(2+D+(2-D)\rho^2)}\, \beta.
	\end{equation}
	
	Finally this gives us the connection
	\begin{equation}
		\beta =\frac{(D-2)}{2(D-1)}\vec\rho\cdot\textbf{K}\vec p.
	\end{equation}

	
	\section{Dyadic Matrix}
	\label{app3}
	To calculate the order parameter in Eq.(\ref{paramAl}), we must perform the integral of the dyadic matrix, which is defined by
	\begin{equation}
		[\hat r\hat r^\top]=\begin{bmatrix}
			\prod_{i=1}^{D-1}\sin^2\theta_i & \prod_{i=2}^{D-1}\sin^2\theta_i\sin{\theta_1}\cos{\theta_1} & \ldots & \sin{\theta_{D-1}}\cos{\theta_{D-1}}\cos{\theta_{D-2}} \\
			\vdots       & \prod_{i=2}^{D-1}\sin^2\theta_i\cos^2\theta_1 &  &  \\
			&  &  \ddots & \\
			\sin{\theta_{D-1}}\cos{\theta_{D-1}}\cos{\theta_{D-2}}& \ldots &  &  \cos^2\theta_{D-1}.
		\end{bmatrix}
	\end{equation}
	
	First notice that all the elements outside the diagonal satisfy 
	\begin{equation}
		\int [\hat r\hat r^\top]_{ij}\, d^DS = 0, \quad \textrm{for}\quad i \neq j .
	\end{equation}
	
	Now, for the terms of the diagonal we have
	\begin{equation}
		\int [\hat r\hat r^\top]_{jj}\, d^DS = \int \prod_{i=j}^{D-1}\sin^2\theta_i\cos^2\theta_{j-1} \sin^{D-2}\theta_{D-1}\sin^{D-3}\theta_{D-2}\ldots\sin{\theta_2}\,d\theta_{D-1}\ldots d\theta_1.
	\end{equation}
	
	\textbf{Proposition:} Let the integral of the dyadic matrix be defined as above, then
	\begin{equation}
		\int [\hat r\hat r^\top]_{jj}\, d^DS = \frac{\mathcal{A}_D}{D}\textbf{1}.
	\end{equation}
	
	\textbf{Proof:} Let's proceed by induction.
	(i) For D=2 \qquad $\int_0^{2\pi}\sin^2\theta d\theta = \int_0^{2\pi}\sin^2\theta d\theta= \pi = \frac{2\pi}{2}=\frac{\mathcal{A}_2}{2}$;
	(ii) Now we suppose that for dimension $D-1$
	\begin{equation}
		\int [\hat r\hat r^\top]_{jj}\, d^DS = \int \prod_{i=j}^{D-2}\sin^2\theta_i\cos^2\theta_{j-1} \sin^{D-3}\theta_{D-2}\ldots\sin{\theta_2}\,d\theta_{D-2}\ldots d\theta_1 = \frac{\mathcal{A}_{D-1}}{D-1}.
	\end{equation}
	
	Then, for dimension D, 
	\begin{equation}
		\int [\hat r\hat r^\top]_{jj}\, d^DS = \int \sin^2\theta_{D-1}\sin^{D-2}\theta_{D-1}d\theta_{D-1}\frac{\mathcal{A}_{D-1}}{D-1} = \frac{\mathcal{A}_{D-1}}{D-1}\int \sin^{D}\theta_{D-1}d\theta_{D-1} .
	\end{equation}
	
	From the reduction formula we have
	\begin{equation}
		\int_a^b \sin^{D}x dx = -\frac{1}{D}\sin^{D-1}x \cos{x}|_a^b + \frac{D-1}{D} \int_a^b \sin^{D-2}x dx.
	\end{equation}
	
	For $a=0$ and $b=\pi$
	\begin{equation}
		\int_a^b \sin^{D}x dx = -\frac{(D-1)(D-3)}{D(D-2)} \int_0^\pi \sin^{D-4}x dx.
	\end{equation}
	
	Repeating the reduction formula we end up with 
	\begin{equation}
		\int_a^b \sin^{D}x dx =
		\begin{cases}
			\frac{(D-1)(D-3)\ldots 1}{D(D-2)\ldots 2}\,\pi, \qquad \textrm{if D is even};\\    
			\frac{(D-1)(D-3)\ldots 2}{D(D-2)\ldots 3}\,2 , \qquad \textrm{if D is odd}.
		\end{cases}
	\end{equation}
	
	So if D is even
	\begin{equation}
		\begin{split}
			\int [\hat r\hat r^\top]_{jj}\, d^DS &=  \frac{\mathcal{A}_{D-1}}{D-1}\frac{(D-1)(D-3)\ldots 1}{D(D-2)\ldots 2}\,\pi\\ &=\frac{\mathcal{A}_{D-1}}{D-1}\frac{(D-2)!}{[2^{\frac{D-2}{2}(\frac{D-2}{2})!}]^2}\,\pi\\ &=\frac{\mathcal{A}_{D-1}}{D}\frac{\Gamma(D-1)\pi}{2^{D-2}[\Gamma(\frac{D}{2})]^2}.
		\end{split}
	\end{equation}
	
	Notice that from Eq.(\ref{AreaSphere})
	\begin{equation}
		\frac{\mathcal{A}_D}{\mathcal{A}_{D-1}}=\frac{\Gamma(\frac{D-1}{2})\pi^{\frac{1}{2}}}{\Gamma(\frac{D}{2}} = \frac{\Gamma(D-1)\pi}{2^{D-2}[\Gamma(\frac{D}{2})]^2},
	\end{equation}
	where we applied the Legendre duplication formula $\Gamma(\frac{D-1}{2})=2^{2-D}\pi^{\frac{1}{2}}\frac{\Gamma(D-1)}{\Gamma(\frac{D}{2})}$ and we have the statement of the proposition satisfied for even dimensions.
	
	Now, if D is odd, 
	\begin{equation}
		\begin{split}
			\int [\hat r\hat r^\top]_{jj}\, d^DS &=  \frac{\mathcal{A}_{D-1}}{D-1}\frac{(D-1)(D-3)\ldots 2}{D(D-2)\ldots 3}\,\,2\\ &=\frac{\mathcal{A}_{D-1}}{D}\frac{[2^{\frac{D-3}{2}}(\frac{D-3}{2})!]^2}{(D-2)!}\,\,2\\
			&=\frac{\mathcal{A}_{D-1}}{D}\frac{[2^{D-2}\Gamma(\frac{D-1}{2})]^2}{\Gamma(D-1)}.
		\end{split}
	\end{equation}
	
	Again, applying the Legendre duplication formula
	\begin{equation}
		\int [\hat r\hat r^\top]_{jj}\, d^DS =\frac{\mathcal{A}_{D}}{D}.
	\end{equation}
	
	For now we have that 
	\begin{equation}
		\int [\hat r\hat r^\top]_{jj}\, d^DS = \frac{\mathcal{A}_D}{D}\textbf{1}.
	\end{equation}

	
	\section{Instantaneous synchronization of non-identical oscillators}
	\label{app4}
	
	When $\omega_1=\omega_2=\omega_4=0$, we observe instantaneous synchronization of non-identical oscillators. We show below that the solution $p=1$ is stable for $K >0$ and unstable otherwise, characterizing a discontinuous phase transition. In this case the equilibrium equations simplify to
	\begin{equation}
		\begin{split}
			0 &= -\omega_6\alpha_2 + \omega_5\alpha_3 -Kp\alpha_4\alpha_1\\
			0 &= \omega_6\alpha_1 -\omega_3\alpha_3 -Kp\alpha_4\alpha_2\\
			0 &= -\omega_5\alpha_1 + \omega_3\alpha_2 -Kp\alpha_4\alpha_3\\
			0 &= Kp(1-\alpha_4^2).
		\end{split}
		\label{equilibcomp}
	\end{equation} 
	
	If $p=0$ we obtain
	\begin{equation}
		\begin{split}
			\omega_6\alpha_2 &= \omega_5\alpha_3\\
			\omega_6\alpha_1 &= \omega_3\alpha_3\\
			\omega_5\alpha_1 &= \omega_3\alpha_2\\
			0 &= Kp(1-\alpha_4^2).
		\end{split}
		\label{equilibcomp2}
	\end{equation} 
	Considering that $\alpha_1, \alpha_2$ and $\alpha_3$ satisfy Eq.(\ref{equilibcomp2}) for all $\omega_3, \omega_5$ and $\omega_6$ then, $\alpha_1=\alpha_2=\alpha_3=0$ and $\alpha_4=\pm 1$, which means that $\vec\rho=\pm\hat x_4$.
	
	If $p\neq0$ the solution is also $\alpha_1=\alpha_2=\alpha_3=0$ and $\alpha_4=\pm 1$, and again $\vec\rho=\pm\hat x_4$.
	
	Now we consider the scenario where $\vec{\rho}$ is perpendicular to $\vec{p}$, i.e., $\alpha_4=0$. In this case the components of Eq.(\ref{eqm1f}) are 
	\begin{equation}
		\begin{split}
			0 &= -\omega_6\alpha_2 + \omega_5\alpha_3\\
			0 &= \omega_6\alpha_1 -\omega_3\alpha_3\\
			0 &= -\omega_5\alpha_1 + \omega_3\alpha_2\\
			0 &= Kp(1+\alpha_1^2+ \alpha_2^2 +\alpha_3^2)
		\end{split}
	\end{equation} 
	and the system has no real solution. 
	
	We now study the stability of these solutions. The Jacobian of this system is 
	\begin{equation}
		\mathbf{J}_4 = \left( 
		\begin{array}{cccc}
			-Kp\alpha_4 & -\omega_6 & \omega_5 & 0\\
			\omega_6 & -Kp\alpha_4 & -\omega_3 & 0\\
			-\omega_5 & \omega_3 & -Kp\alpha_4 & 0\\
			0 & 0 & 0 & -2Kp\alpha_4\\
		\end{array}
		\right).
		\label{jacobian}
	\end{equation}
	and its eigenvalues are $\{-2Kp\alpha_4, -Kp\alpha_4, -Kp\alpha_4\pm i\omega\}$, where $\omega = \sqrt{\omega_3^2+\omega_5^2+\omega_6^2}$. 
	This indicates stability for $Kp\alpha_4>0$. Since the ansatz vector is the order parameter of the set of oscillators with the same matrix of natural frequencies $\vec p\cdot\vec\rho>0$ and the solution $ \vec\rho=\hat x_4$ is stable for $K>0$. Also 
	\begin{equation}
		\begin{split}
			\vec p = \int g(\mathbf{W}_4)d\mathbf{W}_4\hat x_4=\hat x_4
		\end{split}
		\label{pstable}
	\end{equation}  
	and $p=1$. For $K<0$, in the other hand, the only stable solution is $p=\rho=0$.
	
	For $D=2d > 4$ the matrix of natural frequencies has $n=d(2d-1)$ independent variables and is given by 
	\begin{equation}
		\mathbf{W}_{2d} = \left( 
		\begin{array}{cccccc}
			0 &  &  &  &  &  \\
			\omega_n & 0 &  &  &  &  \\
			& \omega_{n-1-2(d-1)} & 0 &  &  &  \\
			&  & \omega_{n-1-4(d-1)} & 0 &  &  \\
			\vdots & \vdots & \vdots &  & \ddots &  \\
			\omega_{n-2(d-1)} & -\omega_{n-4(d-1)} & \omega_{n-4(d-1)-(2d-3)} & \hdots & \omega_1 & 0
		\end{array}
		\right).
		\label{wmateven}
	\end{equation}
	where the entries above the diagonal are set to $(W_{2d})_{ij} = - (W_{2d})_{ji}$. Again, without loss of generality, we choose the direction of the order parameter as $\vec p = p\hat x_{D}$ and we write the ansatz vector as $\vec\rho = \alpha_1\hat x_1+ \alpha_2\hat x_2 + \hdots + \alpha_D\hat x_D = \sum_{i=1}^{D}\alpha_i\hat x_i$. 
	
	The conditions at the equilibrium are the same as for $D=4$:
	\begin{itemize}
		\item If $K=0$ then $\rho= p= 0$.
		\item If $K\neq 0$ and $\rho\neq 0$, then $\rho=1$ or $\vec{\rho}$ is perpendicular to $\vec{p}$.
	\end{itemize}
	
	We first consider $\rho=1$. The components of the Eq.(\ref{eqm1f}) are
	\begin{equation}
		\begin{split}
			\dot{\alpha}_1 &= (\mathbf{W}_D\,\vec\rho)_1 - Kp\alpha_D\alpha_1\\
			\dot{\alpha}_2 &= (\mathbf{W}_D\,\vec\rho)_2 - Kp\alpha_D\alpha_2\\
			\vdots\\
			\dot{\alpha}_D &= (\mathbf{W}_D\,\vec\rho)_D +Kp(1-\alpha_D^2).
		\end{split}
	\end{equation}   
	Analytical solutions similar to Eq.(\ref{delta}) can be obtained but are much more complicated. Here we focus on the discontinuous transition where frequencies  $\omega_{n-2(d-1)}= \omega_{n-4(d-1)}= \omega_{n-4(d-1)-(2d-3)}= \hdots = \omega_1 = 0$. 
	Notice that the last equation gives us $\alpha_D=\pm1$. Once the remaining $D-1$ equations form a null system, there are two possibilities. If the determinant is null there are infinite solutions, but if the determinant is not zero, the only solution is $\alpha_1=\alpha_2=\hdots=\alpha_{D-1}=0$. And that is the case. To prove that, we invoke a theorem from section 3 of \cite{li_mathias_1995}, which we state below.
	
	\textbf{Theorem:} If the $D\times D$ complex matrices $A=A^\top$ and $B=-B^\top$ have singular values, $a_1\geq\hdots\geq a_D\geq 0$ and $b_1=b_2\geq b_3=b_4\geq\hdots$ such that $det(A+B)= z$, then
	\begin{equation}
		z\geq\left\{\begin{array}{lr}
			0, & \textrm{if}\quad [a_n, a_1]\cap [b_n, b_1]\neq \oslash ;\\
			|det(iX+Y)|, & \textrm{otherwise}, 
		\end{array}
		\right.
	\end{equation}
	where $X=\sum_{j=1}^D a_j E_{jj}$ and $Y= \sum_{k\leq\frac{n+1}{2}} b_{2k}(E_{2k-1, 2k}-E_{2k, 2k-1})$. Recall that the singular values of X are the non-negative square roots of the eigenvalues of $X^\top X$

	In our case, the remaining $D-1$ system is characterized by a matrix that can be written as the sum of a skew symmetric matrix and a diagonal matrix $-Kp\alpha_D\mathbf{1}$. Notice that, in this case, $a_1=\hdots= a_D$ and that $B^\top B=-B^2$ and, since $B$ is a skew-symmetric matrix, $-B^2$ is a positive semi-definite matrix. Notice that, since the dimension of $B$ is even, it's eigenvalues are pairs of complex numbers and complex conjugate of them. Consequently, since the eigenvectors of $B$ form a basis of the space, the eigenvalues of $-B^2$ (the singular values) are the square of the eigenvalues of $B$, the the inequality $b_1=b_2\geq b_3=b_4\geq\hdots$ is satisfied.

	If $\vec{\rho}$ is perpendicular to $\vec{p}$, then $\alpha_D=0$ then, the system with $\omega_{n-2(d-1)}= \omega_{n-4(d-1)}= \omega_{n-4(d-1)-(2d-3)}= \hdots = \omega_1 = 0$ is given by
	\begin{equation}
		\begin{split}
			0 &= (\mathbf{W}_D\,\vec\rho)_1\\
			0 &= (\mathbf{W}_D\,\vec\rho)_2\\
			\vdots\\
			0 &= Kp(1+ \alpha_1^2+ \alpha_2^2 + \hdots+ \alpha_{D-1}^2).
		\end{split}
	\end{equation} 
	
	This system has no real solution once the first $D-1$ equations form a null system with non-zero determinant, then again $\alpha_1=\alpha_2=\hdots=\alpha_{D-1}=0$, but this does not satisfy the last equation. 
	
	Now we have that for $K=0$, $p=0$ and for $K\neq 0$ $p=1$ or $p=0$. We want to determine which of these are the stable solution for  $K>0$ and $K<0$.
	
	In the case that $K\neq 0$ and $\rho\neq 0$ we have that $\rho=\hat{x}_D$. So we take a perturbed $\vec\rho=\hat{x}_D + \vec\epsilon(t)$ or $\rho=1+\epsilon(t)$. Then, Eq.(\ref{eqm2f}) to the first order gives us 
	\begin{equation}
		\frac{d\epsilon}{dt}= \frac{K}{2}(1-(1+\epsilon)^2)(\hat\rho\cdot\vec p) = \frac{K}{2}(1-(1-2\epsilon +\epsilon^2))(\hat\rho\cdot\vec p) = -K(\hat\rho\cdot\vec p)\epsilon.
		\label{stability}
	\end{equation} 
	
\begin{figure*}[h]
		\centering 
		\includegraphics[scale=0.5]{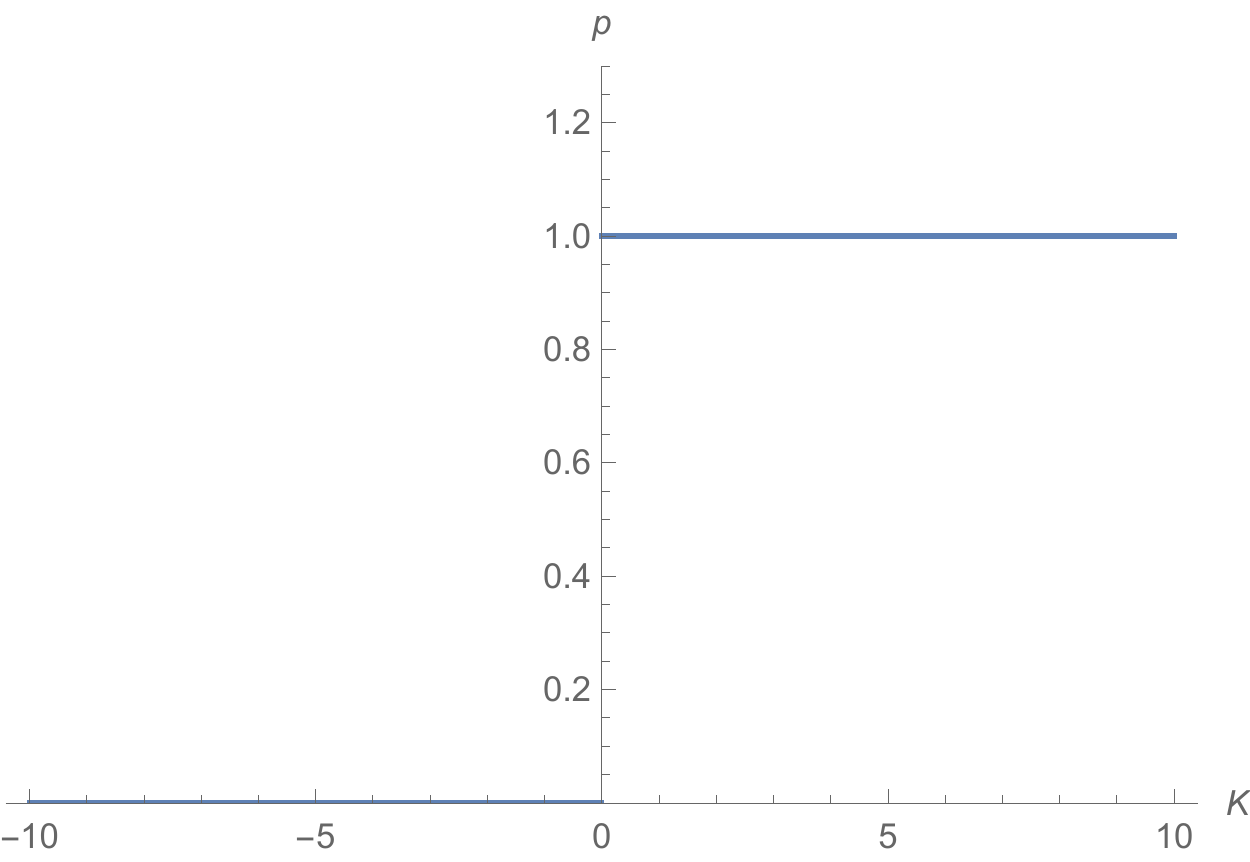} 
		\caption{Stable solution of $p$ as a function of $K$ for the case where $\omega_1=\omega_2=\omega_4=0$. This choice makes the full synchronized state, $p=1$ the stable solution for any $K>0$} 
		\label{stepfunc}
	\end{figure*}

	This means that the stable solution for $K>0$ is the one with $\hat\rho\cdot\vec p>0$, i.e, $p=1$. And for $K<0$ the stable solution is $p=0$, since $\hat\rho\cdot\vec p$ can't be less than zero. This is because $\rho$ is the order parameter of a subset with natural frequency $\omega$, so it cannot be in a direction opposite to $p$.
	We then conclude that the system instantly synchronizes without requiring a scenario of identical frequencies, as we can see in figure \ref{stepfunc}.

\end{document}